# Praseodymium doping effect on the superconducting properties of $FeSe_{0.5}Te_{0.5}$ bulks under ambient and high-pressure growth conditions

Priya Singh[1], Manasa Manasa[1], Mohammad Azam[1], Tatiana Zajarniuk[2], Svitlana Stelmakh[1], Taras Palasyuk[1], Jan Mizeracki[1], Tomasz Cetner[1], Andrzej Morawski[1], Cezariusz Jastrzębski[3], Michał Wierzbicki[3], Shiv J. Singh[1††]

*[1]Institute of High Pressure Physics (IHPP), Polish Academy of Sciences, Sokołowska 29/37, 01-142 Warsaw, Poland*

*[2]Institute of Physics, Polish Academy of Sciences, aleja Lotników 32/46, 02-668 Warsaw, Poland*

*[3]Faculty of Physics, Warsaw University of Technology, Koszykowa 75, 00-662 Warsaw, Poland*

[††]Corresponding author: Shiv J. Singh

Email: sjs@unipress.waw.pl

# Abstract

A series of Pr-doped $FeSe_{0.5}Te_{0.5}$ ($Fe_{1-x}Pr_xSe_{0.5}Te_{0.5}$; $x$ = 0 to 0.3) bulks are prepared by conventional synthesis process at ambient pressure (CSP), and high gas pressure and high temperature synthesis (HP-HTS) methods. These bulks are well characterized by structural and microstructural analysis, Raman spectroscopy, transport, and magnetic measurements. The HP-HTS process of the parent bulks has enhanced the onset transition temperature ($T_c^{onset}$) by 1.5 K and the critical current density ($J_c$) by two orders of magnitude compared to the CSP method. Pr-doped $FeSe_{0.5}Te_{0.5}$ up to 10% doping content prepared, either CSP or HP-HTS, slightly increases the unit cell volume, and high-pressure growth produces an almost pure superconducting phase, which confirms the successful Pr-doping at Fe sites. Raman spectroscopy measurements and DFT calculations suggest the substitution of Pr-atoms in the interlayer spacing of Fe(Se,Te) lattice. High-pressure growth of $Fe_{1-x}Pr_xSe_{0.5}Te_{0.5}$ also makes the sample less dense compared to the parent sample grown by HP-HTS. Transport and magnetic measurements depict that $T_c^{onset}$ is almost unaffected by Pr-doping, whereas $J_c$ of Pr-doped $FeSe_{0.5}Te_{0.5}$ is enhanced by one order of magnitude relative to the parent sample developed by CSP but lower than that of the parent sample grown by HP-HTS. Hence, Pr-doping at Fe sites preserves $T_c^{onset}$ and improves $J_c$ of $FeSe_{0.5}Te_{0.5}$ regardless of the doping contents and growth conditions. These results are promising for the practical application of iron-based superconductors to improve $J_c$ properties without affecting $T_c^{onset}$ through CSP process and congruent with discoveries from other superconductors, like cuprates and $MgB_2$.

# Introduction

The iron-based superconductors (FBS) were discovered in 2008 [1] and became a strong contender for the practical applications due to their high transition temperature ($T_c$) (up to ~58 K) [2], high upper critical field ($H_{c2}$) of 60 T, and ultra-high critical current density ($J_c$) of $10^7$-$10^8$ A/cm$^2$ [3, 4, 5]. After the discovery, more than 100 superconducting compounds are available under this family with a tetragonal layered structure. The first report depicted the superconducting transition of 26 K for F doping of LaFeAsO [1], and the $T_c$ value has reached up to 58 K during the research of the last 15 years [3, 6, 7]. FBS has been recognized as a strong contender for the superconducting magnets, which can be an alternative to the commercially available low-temperature superconducting materials such as NbTi and $Nb_3Sn$ [8]. Furthermore, $MgB_2$ superconductor is limited for low magnetic field response due to its quite low upper critical field and irreversibility field [9], however, FBS materials have a high $T_c$ and high critical magnetic field response compared to $MgB_2$, NbTi, and $Nb_3Sn$, and are more promising materials for high field magnetic applications [6].

FeSe is the simplest family of FBS, belonging to so called 11 family and was discovered in 2008 with an onset $T_c$ of 8 K [7]. Following this pioneering research, the superconducting $T_c$ has been enhanced up to 15 K by Te doping at Se [10] and to 36 K by applied pressure [7, 11]. This system is relatively different from the pnictides, and its superconducting properties are very sensitive to the stoichiometry and applied pressure [7]. Furthermore, no long-range magnetic order is observed, and the superconducting phase is of the orthorhombic or even monoclinic symmetry, where the structural transition from tetragonal to orthorhombic is observed below 110 K. Interestingly, the FeTe compound does not depict superconductivity, but the Te doping at Se sites shows the highest $T_c$ of ~15 K for $FeSe_{0.5}Te_{0.5}$. Several techniques have been employed to improve the superconducting properties of the 11 family, in which chemical doping plays an important role and induces the lattice substitution, strain/stress, and the formation of the secondary phase. Generally, FeSe can exhibit a tetragonal and hexagonal phase at ambient temperature; the tetragonal phase shows the superconductivity. The hexagonal phase is an impurity phase and is not superconducting [12]. However, it's very hard to reduce the hexagonal phase completely during the formation of the tetragonal phase. On the other hand, the presence of secondary phase inclusions with different sizes produced by additions or reactions of dopants enhances the pinning properties, the critical current density, and the improvement of the grain connections [13, 9]. In the case of $MgB_2$ superconductor, the rare earth or rare earth oxide doped samples have shown the superconducting properties without affecting the critical transition temperature, but with a significant improvement in $J_c$ [13]. Almost all high-temperature superconducting materials rely significantly on rare earth metals. For example, Er/Y addition to $Bi_2Sr_2CaCu_2O_{9+\delta}$ did not significantly change the superconducting transition temperature, but increased the transition width and enhanced the critical current properties by improving the flux pinning [14]. In the case of $SmFeAsO_{0.7}F_{0.3}$, the transition temperature and critical current density are improved by the rare earth doping by smaller ($Gd^{3+}$) and larger ($Ce^{3+}$) ions at the $Sm^{3+}$ site [15]. The parent CaFeAsF does not show the superconductivity, but by rare earth element doping into the $(CaF)^+$ layer in this system, superconductivity was observed at 56 K in Nd-doped and at 52 K in Pr-doped samples [16]. Hence, it would be very interesting to explore the rare earth doping effect for the simplest family of FBS, i.e., the 11 family. Last year, our group reported Gd-addition effect on the superconducting properties of Fe(Se,Te) [17] by CSP and HP-HTS, where the onset transition is almost unchanged with Gd-addition but the critical current density is reduced. However, no other studies have been reported for rare earth substitution/doping effects on the superconducting properties of Fe(Se,Te). Hence, it would be worth exploring the rare earth doping effect on the sample quality and properties of Fe(Se,Te) superconductors, and comparing them with the results reported for Gd-addition. This is our main motivation behind this research using CSP and the HP-HTS process.

In this paper, we report the results for a series of Pr-doped $FeSe_{0.5}Te_{0.5}$ bulks ($Fe_{1-x}Pr_xSe_{0.5}Te_{0.5}$; $x$ = 0-0.3) prepared by the conventional and high-pressure growth methods to understand the chemical pressure and applied growth pressure effect. Structural and microstructural analysis, transport, magnetic, and Raman spectroscopy measurements have been performed on these bulks. The increment in the unit cell volume of $Fe_{1-x}Pr_xSe_{0.5}Te_{0.5}$ confirms the successful incorporation of Pr inside the crystal lattice. Interestingly, the hexagonal phase is reduced for a small Pr-doping content ($x \leq 0.1$) through the CSP method, and a pure phase of the tetragonal $FeSe_{0.5}Te_{0.5}$ phase is observed by the high-pressure growth process as cleared from structural analysis. The onset transition temperature is almost the same, and the critical current density is enhanced for these Pr-doped samples prepared by CSP or HP-HTS compared to that of the parent sample ($x$ = 0). High-pressure synthesis improved the parent sample quality with the improvement of $T_c$ by 1.5 K and enhanced the $J_c$ value by two orders of magnitude as compared to the parent sample by CSP. However, Pr-doped Fe(Se,Te) samples prepared by HP-HTS have almost the same onset $T_c$ and $J_c$ values as those prepared by CSP. These results are compared with our previous report based on Gd-added $FeSe_{0.5}Te_{0.5}$ prepared by CSP and HP-HTS [17].

## Experimental details

$Pr_xFe_{1-x}Se_{0.5}Te_{0.5}$ bulks were prepared by conventional two-step solid-state reaction methods with $x$ = 0, 0.01, 0.02, 0.03, 0.05, 0.07, 0.1, 0.2, and 0.3. Fe powder (99.99% purity, Alfa Aesar), Se powder (99.99% purity, Alfa Aesar), Te powder (99.99% purity, Alfa Aesar), and Pr powder (99.99% purity, Alfa Aesar) were used as the starting precursors and mixed according to the stoichiometric formula $Fe_{1-x}Pr_xSe_{0.5}Te_{0.5}$ in an agate mortar and then prepared into the cylindrical disks with a diameter of 11 mm and a thickness of around 2-3 mm by applying hydraulic pressure of 20 bar. These pellets were sealed into an evacuated quartz tube with a vacuum of the order of ~$10^{-2}$-$10^{-3}$ bar, placed inside the furnace, and heated at 600°C for 11 hours. They were then cooled down to room temperature with a slow and controlled cooling rate. After this reaction process, this quartz tube was opened inside the glove box, and the obtained pellets were reground. These powders were pelletized again, having the same diameter as the first step, which was sealed again in the evacuated quartz tube. This sealed tube was heated again at 600°C for 4 hours inside the same furnace. The final disks were blackish and had a same diameter and thickness as those initially. All preparation processes were performed in a highly pure inert gas atmosphere glove box with very low ppm oxygen and moisture levels. All samples were prepared under the same growth conditions and with the same instruments to maintain their reproducibility. Further details on conditions and growth processes can be found in our previous papers [18, 17, 12]. The list of samples and the impurity phase content are presented in Table 1.

To understand the high-pressure growth effect, we have prepared several Pr-doped samples, with Pr content $x$ = 0, 0.01, 0.02, 0.05, and 0.10, by the high gas pressure and high-temperature synthesis (HP-HTS) method. Our HP-HTS technique utilizes a cylindrical pressure chamber and one can apply pressure up to 1.8 GPa. A single- and three-zone furnace which is equipped with this chamber, can create a temperature of up to 1700°C. One can find more details about our HP-HTS elsewhere [19]. After CSP process, the quartz tube was opened inside the glove box, and the obtained pellet was sealed in the Ta tube under the Ar atmosphere through an ARC melter. The sealed Ta tube was placed inside the pressure chamber and applied to the optimized pressure growth conditions: 500 MPa and 600°C for 1 hour through HP-HTS. In our previous study [12], we optimized the growth conditions of Fe(Se,Te) by preparing the bulks $FeSe_{0.5}Te_{0.5}$ in a wide range of applied pressure 0-1.8 GPa and following various growth conditions. We conclude that high-quality Fe(Se,Te) bulks can be prepared under the following conditions: 500 MPa, 600°C, 1 h, and the sample must be sealed into a Ta tube. Based on these findings, we have applied these best conditions to grow the Pr-doped Fe(Se,Te) using the HP-HTS method.

Structural characterization of these samples was performed by X-ray diffraction patterns using an X'Pert PRO, Panalytical diffractometer with filtered Cu–Kα radiation (wavelength: 1.5418 Å, power: 30 mA, 40 kV) and a $PIXcel^{1D}$ position scintillation detector. A measurement profile from 10 to 70° was applied with a very small step of 0.01 deg/min and a slow scan for all prepared samples. The profile analysis was conducted using the ICDD PDF4+ 2021 standard diffraction patterns database and Rigaku's PDXL software. These investigations yielded quantitative values of impurity phases (%) and lattice characteristics for various samples. An examination of the microstructure and elemental mapping was conducted using a Zeiss Ultra Plus field-emission scanning electron microscope with an EDS microanalysis system by Bruker Mod. Quantax 400 and an ultra-fast detector. The magnetic properties of the materials were characterized using a vibrating sample magnetometer (VSM) connected to a Quantum Design PPMS. The measurements were conducted in a temperature range of 5–25 K and a magnetic field of up to 9 T. The magnetic moment was measured under zero-field-cooled (ZFC) and field-cooled conditions (FC) with an applied magnetic field of 50 Oe using a slow temperature scan. The temperature dependence of resistivity was measured using a closed-cycle refrigerator in a zero magnetic field using different current levels (5, 10, and 20 mA), the data were obtained during a gradual warming process.

Two series of polycrystalline samples of pristine $FeSe_{0.5}Te_{0.5}$ and with Pr-doping with the nominal composition $Fe_{1-x}Pr_xSe_{0.5}Te_{0.5}$ ($x$ = 0.05, 0.10) were used for Raman scattering measurements. Samples from the first series were obtained via ambient pressure synthesis, whereas samples of the second series were synthesized at high temperature and pressure conditions. We performed optical measurements on the sample surface immediately after synthesis, without any additional treatment. We first executed a preliminary investigation of the parent $FeSe_{0.5}Te_{0.5}$ at room temperature. It was found that the compound was not stable and was undergoing decomposition and subsequent oxidation. Decomposition was observed by the progressive appearance of elemental Tellurium, whereas the process of oxidation was confirmed by the presence of iron oxide $Fe_2O_3$ (hematite). It is worth mentioning that the spectra of Te, hematite, or their combination are very often mistaken for the characteristic spectra of FeSe/Te-based materials [20, 21, 22, 23]. We also found that degradation was significantly accelerated at higher laser power, probably due to local heating of the sample. To suppress degradation processes, we performed Raman scattering measurements at low-temperatures in an oxygen-free atmosphere. For that purpose, samples were loaded inside a temperature-control microscope stage (Linkam DSC600) that was simultaneously flushed with gaseous nitrogen. Raman spectra were collected at a temperature of 113 K (-160ºC). Laser power was attenuated to 0.08 mW to prevent local heating of the sample by the tightly focused laser beam. In order to check the effect of local heating during a low-temperature study, selected measurements were conducted at the same probing point on the sample using a higher laser power of 0.8 mW.

Raman scattering measurements were performed using a LabRam ARAMIS (Horiba Jobin Yvon) spectrometer. Samples were excited by visible light from an ion He-Ne laser with a 632.8 nm wavelength. Incident light was focused by a 50x objective with NA 0.55 (Olympus ULWD) to a spot of < 5 μm on a probed sample. The same objective was used for collecting backscattered light that was further dispersed by a diffraction grating of 2400 l/mm and further registered by a charge-coupled device (CCD), yielding Stokes part of Raman shifts in the range from 65 to 300 $cm^{-1}$. The acquisition time of a single accumulation was set to 600 s (10 min). Signals in the measured spectra were deconvoluted by peaks of Lorentzian shape. As a consequence of the polycrystalline nature of the samples, the collected spectra showed a sizable scatter as regards the intensity of detected peaks, depending on a point across the sample being examined. These observations might be ascribed to different grain orientations. Therefore, we did not perform an analysis of doping-induced effects on peak intensity.

The obtained phonon properties of pure and Pr-doped $FeSe_{0.5}Te_{0.5}$ samples were further assessed by density functional theory (DFT) calculations. Internal parameters of structure geometries, atomic forces and Raman frequencies were determined by the all-electron full-potential linearized augmented plane-wave (LAPW) program WIEN2k [24]. The generalized gradient approximation (GGA) to the exchange-correlation functional was applied using the Perdew, Burke, and Ernzerhof (PBE) form [25]. Plane-wave cut-off parameter $Rk_{max}$ was set to 9, 2000 k-points in the first Brillouin zone were selected, and 6 nonspherical matrix elements for large spheres (LVNS). Values of $Rk_{max}$ and k-points were determined by a set of convergence tests, with forces converged to less than 0.1 Ry/Bohr. The rest of the initialization parameters were set at default values. To facilitate comparison with experimental values of Raman frequencies, lattice constants for all tetragonal crystal structures were selected as: $a$ =

3.7996 Å, $c$ = 5.9895 Å, which correspond to the case of $FeSe_{0.44}Te_{0.56}$ crystal analyzed in [26]. As it is well known that optimization of volume and $c/a$ ratio of crystal cells by DFT calculations may lead to sizeable discrepancies with experimental values of lattice constants [27], therefore we selected the single set of experimental baseline crystal structure parameters to investigate the influence of chemical composition on Raman shifts. In all cases, internal free parameters of Wyckoff positions were optimized by minimizing the internal forces with the SCF convergence criterion of 0.1 mRy/Bohr. The frequencies of Raman modes were determined by the frozen-phonon method. Symmetric Raman-active modes were determined by group theoretical method [28]. Force constants were obtained by the first derivative of 5-point cubic interpolation of force vs. atom displacement relationship.

## Results and discussion

Powder XRD patterns of $Pr_xFe_{1-x}Se_{0.5}Te_{0.5}$ samples prepared by the conventional synthesis process are shown in Figure 1(a). The parent compound $FeSe_{0.5}Te_{0.5}$ has the main tetragonal phase with a space group of *P4/nmm*, and the hexagonal phase around 5-6% is also observed as an impurity phase, like in the previous reports [29, 7, 12]. Interestingly, a small amount of Pr-doping at Fe sites reduces the hexagonal phase (~2-3%), and a tetragonal phase is observed as the main phase. A further increase in Pr-doping, i.e., $x$ = 0.1, leads to a slightly higher amount of hexagonal phase, and an extra phase of $Fe_5Te_4$ is also observed in contrast to the low amount of Pr-doped and parent samples. Large amount of doping, like $x$ = 0.2 and 0.3, rapidly enhanced the hexagonal phase and clearly revealed the PrSe and $Fe_5Te_4$ phases. For example, the $x$ = 0.3 sample has more than 40% of hexagonal phase. The amount of the impurity phases for the different samples is listed in Table 1. It suggests that up to 10% Pr-doping at Fe sites supports the tetragonal phase formation, and further increases in Pr-doping enhance the hexagonal phase very rapidly. The larger ionic radius of Pr (113 pm) compared to Fe (63 pm) ion causes a slight shift in the main peak position 101, suggesting small changes in the lattice parameters.

Figure 1(b) depicts the XRD data for various $Pr_xFe_{1-x}Se_{0.5}Te_{0.5}$ bulks processed by HP-HTS. The sample $x$ = 0_HIP has almost the same XRD pattern as that of the parent sample processed by CSP. Pr-doped samples, i.e., $x$ = 0.01_HIP, 0.02_HIP, and 0.05_HIP have reduced the hexagonal phase, and an almost clean superconducting tetragonal phase is observed, which is almost the same as that of these samples processed by CSP. A large amount of Pr substitution, such as $x$ = 0.1_HIP, has increased the hexagonal phase again, with content slightly higher than that for the samples processed by CSP (Table 1). Hence, the appearance of the hexagonal phase has almost the same trend for $Pr_xFe_{1-x}Se_{0.5}Te_{0.5}$, either processed by CSP or HP-HTS. It suggests that the hexagonal phase is reduced and the purity of samples is increased for the low amount of Pr-doping i.e., $x\leq0.1$. The main 101 peak position of various HP-HTS processed Pr-doped samples confirms a slight shift compared to the parent sample, i.e., *x* = 0_HIP. This behaviour is the same as that observed for the samples processed by CSP.

The calculated lattice parameters '*a*' and '*c*', and unit cell volume '*V*' are plotted in Figures 1(c), (d) and (e) for various Pr-doped samples (*x*) prepared by the CSP and HP-HTS processes. A small enhancement of the lattice parameters is clearly observed with Pr dopants, for the samples prepared either with CSP or HP-HTS. This agrees with the results presented in Figures 1(a)-(b). The variation in the lattice parameter '*c*' is more visible than that of the lattice '*a*' with respect to Pr-doping. To be clearer about the rare earth doping effects, we have also included the lattice parameters '*a*' and '*c*' for Gd-addition to $FeSe_{0.5}Te_{0.5}$ bulks from the previous report [17]. Due to the larger size of Gd (94 pm) compared to Fe (63 pm), the enhancement of the lattice parameters and unit cell volume is also observed, as depicted in Figures 1(c)-1(e), which is well in agreement with both Gd-addition and Pr-doped samples prepared by CSP and HP-HTS. Interestingly, more than 10% of Pr-doped samples have almost constant lattice parameters, which suggests that Pr-dopants are not entered inside the tetragonal lattice. Due to this, an impurity phase related to Pr, i.e., PrSe is also observed in a large amount for the samples *x* = 0.2 and 0.3, and its presence confirms that Pr is not entering the superconducting tetragonal phase with an agreement of the calculated lattice parameters. These behaviours are well agreed with those observed for Gd-additions.

Typical Raman spectra collected in our study are shown in Figure 2 for the parent $FeSe_{0.5}Te_{0.5}$ and 5% Pr-doped $FeSe_{0.5}Te_{0.5}$ at room temperature (300 K) and low-temperature (113 K), respectively. As mentioned above, samples were excited using three levels of laser power: high (4 mW), medium (0.8 mW), and low (0.08 mW) to check the stability of the sample with respect to decomposition and oxidation. Raman spectra collected at room temperature with high laser power, shown in Figure 2(a) confirm the presence of elemental Tellurium [30] as well as iron oxide $Fe_2O_3$ (hematite). At low-temperature and medium laser power, as depicted in Figure 2(b), the elemental Te spectrum reveals the presence of signals related to chalcogen Se/Te ($A_{1g}$) and iron Fe ($B_{1g}$) out-of-plane lattice modes, at ca. 154 $cm^{-1}$ and ca. 203 $cm^{-1}$, respectively. These values are consistent with earlier observations at low-temperature conditions for single crystalline materials with nominal compositions of $Fe_{0.95}Se_{0.56}Te_{0.44}$ [31] and $FeSe_{0.5}Te_{0.5}$ [32]. We note that in our measurements, the Se/Te phonon frequency is ca. 6 $cm^{-1}$ lower than the reported value for Fe(Se,Te) single crystals [20, 21] at $T \approx 113$ K. It has been shown in Ref. [31], a small softening of Se/Te ($A_{1g}$) mode can be induced by the excess Fe in the sample. For Pr-doped samples, *i.e.*, $Fe_{1-x}Pr_xSe_{0.5}Te_{0.5}$ ($x$ = 0.05, 0.10), no major changes in the Raman spectra were detected which suggest the stability of crystal structure as well as unchanged chemical bonding upon addition of praseodymium, as shown in Figures 2(c) and 2(d). Moreover, Raman shifts associated with vibrations of chalcogen and iron sublattices were found to show close values for parent and Pr-doped samples (irrespective of the amount of doping), as depicted in Figure 2(d). The observed insensitivity of phonon frequencies for Pr-doped samples may indicate that the addition of Pr does not have a considerable effect on the dynamics of iron sublattice.

According to our DFT calculations, one should expect a significant lowering of phonon frequencies related to metal sublattice in the case of the effective substitution of iron by heavier praseodymium atoms (Figure S1). On the other hand, according to the above-discussed XRD measurements, the volume of the $FeSe_{0.5}Te_{0.5}$ tetragonal unit cell is considerably increased with the addition of Pr. Furthermore, the crystal lattice expands mainly along the "*c*" direction, while the expansion along the "*a*" direction is one order of magnitude lower, as shown in Figure 1 (c) and (d), respectively. It suggests that Pr atoms fill the interlayer spacing, most likely between adjacent Ch–Fe–Ch (Ch = Se/Te with 0.5 occupation) trilayers. Hence, such intercalation by Pr atoms should have a rather minor influence on Fe – Ch bonding, which can explain the almost unperturbed Raman shift found in our Raman scattering measurements for different amounts of Pr-doping. The interlayer intercalation mechanism of doping suggested here for Pr is different in comparison to the way Cu doping is introduced into SmFeAs(O,F) crystal. We have recently shown that doping of Cu atoms effectively substitutes Fe atoms within the Fe–As layer, and that has an immediate effect on Raman frequency of Fe (B1g) and As (A1g) phonons [33]. Further on, unlike the present case of Pr, the intralayer mechanism of doping was evidenced by the nearly constant character of the lattice parameter "*c*" upon increasing amounts of Cu dopants, whereas the lattice parameter "*a*" experienced a small but well-determined expansion. *RE* monochalcogenide compounds are known to form cubic crystal lattices rather than tetragonal crystals surrounding them [34, 35]. Considering the assumed interlayer positioning of Pr atoms (with Se/Te layers as the closest neighbors), this may explain why, for higher Pr-doping (greater than 10%), we observed the formation of PrSe/Te compounds.

To understand the actual contents and distribution of the constituent elements in these samples, the elemental mapping, and the energy dispersive X-ray (EDAX) were performed, which are shown in Figures 3 and 4, whereas other samples are depicted in Figure S2. The actual concentration of various elements in various Pr-doped samples is listed in Table 2, estimated from the EDAX analysis. The parent compound ($x = 0$) prepared by CSP depicts the homogeneous distribution of Fe, Se, and Te as provided in Figure 3(i). The parent sample prepared by high pressure has almost the same homogeneity of the constituent elements as that of parent samples synthesized by CSP, as shown in Figure 4(i). The samples with a small amount of Pr substitution, such as $x = 0.01$ (Figure S2(i)) and 0.02, have a non-homogeneous distribution of Pr element; however, other elements Fe, Se, and Te have an almost homogeneous distribution, as illustrated in Figure 3(ii). The same kind of mapping is also observed for the samples processed by HP-HTS, as illustrated in Figure 4(ii). Further increase of Pr-dopants improves the homogeneous distribution of Pr with other elements, as depicted for $x = 0.05$ in Fig. 3(iii). However, the growth of this sample by HP-HTS promotes the formation of FeTe and PrSe impurity phases and enhances the inhomogeneity of the constituent elements, as shown in Figure 4(iii) for $x = 0.05$_HIP. The inhomogeneous distribution of the constituent elements is observed below for $x = 0.1$ prepared by CSP, but the HP-HTS process supports the phase formation of the hexagonal and PrSe phases rather than the tetragonal phase. These results are almost the same as observed from XRD. The impurity phase PrSe starts to be observed very clearly for the sample $x = 0.1$, as depicted in Figure 3(iv), and Fe and Te rich regions are observed, which confirm the presence of hexagonal phase $Fe_5Te_4$. The similar results are for the sample $x = 0.1$_HIP prepared by HP-HTS and the XRD data analysis agrees well with the findings. Further increase of Pr content enhances the hexagonal ($Fe_7Se_8$ or $Fe_5Te_4$) and PrSe phase formation very rapidly, as suggested by the mapping depicted in Figure 3(v) for $x = 0.3$, supporting XRD data analysis. The entire sample exhibits numerous inhomogeneities of its constituent elements, suggesting that Pr-doping is only suitable for less than 10%. Due to the large area of inhomogeneity for the sample $x = 0.1$_HIP, we have not prepared any samples with higher Pr dopants ($x > 0.1$) by the HP-HTS method.

To analyse the microstructural properties, the backscattered electron (BSE) images are shown in Figures 5 and 6 for some selected samples prepared by CSP and HP-HTS, respectively (whereas other sample images are depicted in Figure S3). The samples were polished with different grades of emery paper inside the glove box without using oil or any lubricant. Black, white, and grey contrasts are observed corresponding to pores or PrSe, hexagonal ($Fe_7Se_8$), and $Pr_xFe_{1-x}Se_{0.5}Te_{0.5}$ phases, respectively. Parent samples prepared by CSP and HP-HTS are depicted in Figures 5(a)-(c) and 6(a)-(c). In both cases, the parent $x = 0$ and $x$ = 0_HIP have almost a clean phase in both cases, but the sample $x = 0$ has many grain boundaries and many nanopores do exist (Figure 5(a)-(c)). The HP-HTS processed sample, *i.e.*, $x$ = 0_HIP, has reduced the number of pores, and many well-connected grain boundaries do exist. It seems that the HP-HTS-processed sample has improved the density by reducing the micro- and nanopores present in the parent $FeSe_{0.5}Te_{0.5}$ obtained by CSP. The sample $x = 0.01$ has more pores than the parent ($x = 0$), and a very small white area could be related to the hexagonal phase, as shown in the supplementary Figure S3. Pr-doped sample $x = 0.02$ exhibits a large whitish area that could be related to the hexagonal phase, and the rest of the sample looks homogeneous. The high-pressure grown sample $x$ = 0.02_HIP is more compact than the sample $x = 0.02$ processed by CSP, but the hexagonal phase is observed as a white contrast. The further increase of Pr contents increases the hexagonal phase with many pores. Interestingly, the sample $x = 0.05$ has a very tiny hexagonal phase and many pores (Figure 5(g)-(i)), but by applying growth pressure, this sample has a larger area of hexagonal phase and improved the compactness of the sample, as shown in Figure 6(g)-(i). With a further increase of Pr contents, the sample $x = 0.07$ also has many white areas, which suggests that the hexagonal phase has increased, but the sample has reduced its compactness as compared with the parent sample (Figure S2 (g)-(i)). The high Pr-doped sample prepared by CSP or HP-HTS, i.e., $x = 0.1$ and $x$ = 0.1_HIP has a large area of hexagonal phase but looks more compact than the lower Pr-doped samples, as depicted in Figure 5(j)-(l) and Figure 6(j)-(l). Further increases of Pr dopants, i.e., $x = 0.2$ (Figure S3) and 0.3, have increased the hexagonal phase very rapidly, and many white areas are observed inside the samples (Figure 6(m)-(o)), which is well agreed with XRD data analysis. These analyses suggest that samples with Pr-doping up to 7% have a very tiny amount of hexagonal phase compared to the parent sample ($x = 0$ and 0_HIP) obtained by either CSP or HP-HTS processes. However, higher Pr dopants ($x$ >0.1) enhance the hexagonal and PrSe phases rapidly in both CSP and HP-HTS. These analyses suggest that Pr-doped samples prepared either CSP or HP-HTS have less density than that of the parent sample ($x = 0$ and 0_HIP).

The temperature dependence of the resistivity of $Pr_xFe_{1-x}Se_{0.5}Te_{0.5}$ samples prepared by CSP is depicted in Figure 7(a)-(d) up to room temperature. The resistivity of the parent $FeSe_{0.5}Te_{0.5}$ sample decreased from room temperature to a low-temperature region, and a large anomaly was observed below 110 K due to the structural phase transition. A small amount of Pr-doping increases the resistivity in the whole temperature range for $x = 0.01$, 0.02, and 0.03 as shown in Figure 7(a). Further increase of Pr, i.e., $x = 0.05$ and 0.07 again enhance the normal state resistivity value rapidly. Interestingly, the 10% Pr-doping, i.e., the sample $x = 0.1$, has a slightly lower resistivity compared to the sample $x = 0.07$, which is not in a systematic trend compared to the low amount of Pr-doping, and it could be possible due to the presence of the impurity phases, many pores, and the less compactness of the sample ($x = 0.07$) as observed in mapping and BSE images. Pr-doping increases the resistivity in the region from 300 K to 150 K, and then the resistivity decreases and superconductivity is observed below 20 K for $x \leq 0.1$. The enhancement of the resistivity is more visible in the higher Pr-doped samples, as observed in Figure 7(a). Further increase of Pr-doping, i.e., $x = 0.2$ and 0.3 has increased the normal state resistivity value at the temperature range from 300 to 20 K just before the superconducting transition, as depicted in Figure 7(b). The almost 50% increase in resistivity value in the low-temperature range (~20 K) compared to that of the resistivity value at the room temperature is observed. This indicates an insulating behaviour due to a large amount of Pr-doping, as clearly observed for the samples $x = 0.2$ and 0.3. However, the onset of the superconducting transition is observed for these samples around 12-13 K, but the zero resistivity is not observed, as shown in the inset of Figure 7(b). It suggests the presence of a non-superconducting phase, which is well agreed upon with structural and microstructural analysis. The anomalous increase of resistivity as a function of decreasing temperature, especially below 100 K as depicted in Figure 7(b), can be due to either weak localization or the Kondo effect [36], as reported for other FBS such as Ni-doped LaOFeAs [37], Co-doped LaFeAsO [38] and Rh-doped NdFeAsO [39]. These samples ($x = 0.2$ and 0.3) also have a large amount of hexagonal phase (35% for $x = 0.2$ and 45% for $x = 0.3$; Table 1) compared to other samples. It has been reported that the hexagonal phase ($Fe_7Se_8$) is antiferromagnetic [40] and exhibits a metal–insulator transition at temperatures below 100 K, suggesting that locked spins at the grain boundaries contribute to electron localization, leading to this transition [41]. Hence, the observed enhancement of resistivity for these Pr-doped bulks below 100 K may be attributed to the significant presence of the hexagonal phase, which likely serves as a source of localization.

The low-temperature resistivity behaviour of the Pr-doped samples up to 10% substitutions is illustrated in Figure 7(c) below 16 K. The parent sample ($x = 0$) has a transition temperature of 14.8 K with a broader transition of ~3.1 K. A small amount of Pr-doping, i.e. $x = 0.01$, has slightly reduced the onset $T_c$ to 13 K, which is ~1.8 K less than the parent sample ($x = 0$). For further increase of Pr dopants, i.e., $x > 0.01$, samples have almost the same onset $T_c$ value of ~14 K, which is almost constant even up to 10% of doping contents. The most noticeable is that Pr-doped samples have reduced the transition broadening by about 1 K, which is almost 2 K lower than that of the parent compound. However, samples with a very large amount of Pr substitutions, i.e., $x = 0.2$ and 0.3, have an onset $T_c$ of 12-13 K, which is slightly lower than that of the samples with $x \leq 0.1$, but no zero resistivity is observed down to the lowest measured temperature of 7 K (the inset of Figure 7(b)). Pr-doping does not affect too much on the onset transition temperature but improves the sharpness of the transition up to 10% of doping.

In order to understand the grain connections of these CSP-prepared samples, we have evaluated the resistivity's temperature dependence at different currents ($I = 5$, 10, and 20 mA). The onset transition temperature ($T_c^{onset}$) is associated with the individual grain effect, also known as the intragrain effect, whereas the offset transition temperature ($T_c^{offset}$) is represented by the grain connections, also known as the intergrain effect. These effects can be comprehended using resistivity measurements at different applied currents. Figure 7(d) shows the resistivity characteristics at low-temperatures for samples subjected to three different currents: 5, 10, and 20 mA, to study the behaviour of grains and grain connections. The onset and offset transition temperatures of the parent sample prepared by CSP are not sensitive to the various currents, which suggests good grain connections. The sample $x = 0.02$ has shown the sensitivity of offset transition temperature having almost constant the onset $T_c$ for the various currents very clearly, which could be due to the inhomogeneity of the sample as observed from the mapping. Further increases in Pr-doping, such as $x = 0.05$, have almost constant the onset Tc, whereas a slight lowering the offset transition is observed with the increasing currents, which seems slightly more observable than that of the parent sample. However, the high Pr-doped samples, such as $x = 0.1$, the offset $T_c$ is reduced rapidly to a lower temperature under various applied currents. Overall, Pr-doped samples processed by CSP have depicted almost the same onset $T_c$ and a shift of the offset $T_c$ to lower temperature with the various currents, i.e, the enhancement of transition width ($\Delta T$), which suggests the reduced grain connections compared to the parent sample ($x = 0$) and supports the presence of the impurity phases as discussed above with structural and microstructural analysis.

The resistivity behaviour of the samples prepared by HP-HTS is shown in Figures 8(a) and 8(b), with a temperature variation up to 300 K. The HP-HTS processed parent sample, i.e., $x$ = 0_HIP, has a lower resistivity and slightly higher $T_c$ value than that of the parent sample ($x = 0$) processed by CSP in the whole temperature region. The sample $x$ = 0.02_HIP has a huge enhancement of the resistivity due to HP-HTS compared to the sample $x = 0.02$, which is almost 10 times higher than that of the sample $x = 0.02$. This clearly suggests the presence of the impurity phase and the reduced grain connections. However, the onset $T_c$ value and transition width of the sample $x = 0.02$_HIP are the same as those of the sample $x = 0.02$, even in the presence of many impurity phases, as shown in the inset of Figure 8(a). The same broadening of $x = 0.02$_HIP is observed due to the compressed sample and high-pressure growth effects, which play a dominant role compared to that of the impurity phase. Figure 8(b) depicts the resistivity behaviours of the sample $x = 0.01$_HIP, 0.05_HIP, and 0.10_HIP. These samples have a very high resistivity value compared to the parent, and $x = 0.02$, 0.02_HIP. Interestingly, the onset $T_c$ is the same for these samples as that of the samples processed by CSP, but zero resistivity has not been reached for these samples. To confirm and recheck the resistivity data of HP-HTS grown samples, measurements were carried out on different samples of the same batch of $x = 0.05$_HIP and $x = 0.01$_HIP and are shown in Supplementary (Figure S4). It could be due to the inhomogeneity of these samples and the presence of the impurity phases, which are enhanced by the high-pressure growth process, which well agrees with microstructural analysis. The intergranular and intragranular behaviors of the polycrystalline samples play a crucial role in determining the characteristics of FBS [42, 43, 44]. The occurrence of zero resistivity is associated with the connections between grains, which can be influenced by factors such as the presence of impurity phases among the superconducting grains and the density of the sample, among others. When the samples are dense and exhibit well-defined and clear grain connections, the superconducting transition width ($\Delta T$) will be more pronounced. As the Pr-doping contents increase, there is a notable increase in the impurity phases, particularly the non-superconducting hexagonal phase (Table 1). The existence of these impurity phases influences the behaviors of both the inter- and intragranular regions by covering the grain and grain boundaries, as reported for thick perovskite FBS [45]. As a result, we did not observe zero resistance for higher Pr-doped samples such as $x = 0.2$ and 0.3. Furthermore, one can also note that the temperature dependence of the resistivity behavior for all these bulks is measured up to 7 K. It could be possible that measuring the temperature dependence of the resistivity down to 2 K may reveal zero resistivity for these samples ($x$ = 0.3, 0.2, 0.01_HIP, 0.05_HIP, 0.1_HIP) accompanied by a very broad transition.

The resistivity behaviour of Pr-doped Fe(Se,Te) is depicted in Figure 8(c) for different currents. The parent $x = 0$_HIP has no broadening with respect to the applied current $I = 5$, 10 and 20 mA compared to the parent sample $x = 0$. It suggests that the parent sample ($x = 0$_HIP) has better grain connections than the sample processed by CSP ($x = 0$). The onset and offset $T_c$ of the samples with low amounts of Pr-doping are shifted to low-temperature by applying the various currents as observed for $x = 0.02$_HIP. It suggests that both the intragrain and intergrain connectivity of Pr-doped samples are affected by HP-HTS process. The suppression of the superconducting transition of $x = 0.02$_HIP is observed due to the increased applied current from 5 mA to 20 mA in zero magnetic field. It suggests that higher bias currents increase the self-field in the superconducting grains, and the dynamics of vortices can be generated by the high bias current [46], [47]. The penetration of vortices along the grain boundaries suppresses the superconducting transition with high applied current, as observed for the sample $x =$ 0.02_HIP (Figure 8(c)). It should be noted that the detailed mechanisms by which a bias current produces magnetic vortices in the absence of a magnetic field remain unclear [48] . However, potential explanations involve "effective" magnetic vortices resulting from non-uniform current densities due to the inherent inhomogeneity of the samples, as well as the formation of magnetic vortex pairs through straight current paths [49]. Since the other HP-HTS process samples have not shown the offset transition temperature, the resistivity in the presence of various currents was not measured. This behaviour also confirms that these samples under HP-HTS process have non-superconducting phase and reduce the density of superconducting phase due to Pr substitution, as observed from the microstructural analysis.

To confirm the Meissner effect of these samples, zero-field-cooling (ZFC) and field-cooled (FC) magnetization have been performed for some selected samples, i.e., $x = 0$, 0.01, 0.02, 0.05, and 0.1 at a magnetic field of 50 Oe. To make comparative studies of these samples, ZFC and FC are shown in Figure 9(a). The parent compound ($x = 0$) has an onset $T_c$ of 14 K, whereas other samples prepared by CSP have almost the same onset $T_c$ value of 12-13 K. All samples have a single-step transition, suggesting that the intergranular properties of these bulks are comparable to those reported for other FBS families [44]. One can note that the parent sample has a broader transition, which is improved for Pr-doped samples. The obtained results are in agreement with those discussed above for transport measurements. Figure 9(b) illustrates the ZFC and FC behaviour of the samples prepared by HP-HTS, i.e., for $x = 0$_HIP, 0.01_HIP, 0.02_HIP, and 0.1_HIP. The 1% Pr-doped sample, i.e., $x = 0.01$_HIP, has a much lower onset $T_c$ than that of the sample $x = 0.01$, whereas $x = 0.02$_HIP has almost the same onset $T_c$ value of 13 as that of $x = 0.02$. ZFC and FC behaviours for a very small amount of Pr-doping have suggested the inhomogeneous distribution of Pr contents through the HP-HTS process, as observed from the microstructural analysis. However, the sample with a large amount of Pr-doping, i.e., $x = 0.1$_HIP, has an onset $T_c$ of 10 K, which is lower than that observed for the resistivity measurements. It suggests that a large amount of Pr-doped sample has a lower superconducting volume fraction, which might be due to the presence of impurity phases. However, the magnetic susceptibility behaviour of all samples processed either by HP-HTS or CSP has almost the same behaviour.

The critical current density plays an important role in the application of a superconductor. To calculate the $J_c$ value, the magnetic hysteresis loops at 7 K were measured for $x = 0$, 0.01, 0.02, 0.05, and 0.1, and $x = 0$_HIP, 0.02_HIP. The Pr-doped sample has ferromagnetic contributions due to the presence of a very tiny amount of iron or a paramagnetic phase due to the presence of impurity phases, so we have subtracted the normal state magnetization, i.e., *M-H* loop recorded at 22 K. These loops are shown in the inset of the figure 9(c) for $x = 0.02$ and 0.02_HIP. The parent sample has slightly different behaviour than that of Pr-doped samples. Using hysteresis loop width $\Delta m$, we can calculate the critical current density using the Bean model [50] and the formula $J_c = 20\Delta m / V_s a(1 - a/3b)$, where $V_s$ is the sample volume, and '*a*' and '*b*' are the length of the sample ($a<b$). The calculated critical current density for the sample $x = 0$, 0_HIP, 0.01, 0.01_HIP, 0.02, 0.02_HIP, 0.05, and 0.1 is shown in Figure 9(c). The parent sample $x = 0$ has a $J_c$ of the order of $10^2$ A/cm$^2$ at 0 T, which is reduced rapidly to $10^1$ A/cm$^2$ at 1 T. This value is almost constant up to 9 T magnetic field. The application of the HP-HTS process enhances the $J_c$ value of the parent compound ($x = 0$_HIP) by around two orders of magnitude, and its behaviours are almost the same as those of the sample ($x = 0$). A small amount of Pr-doping, i.e., $x = 0.01$ and 0.02, has improved the $J_c$ value by almost one order of magnitude compared to that of the parent compound, $x = 0$, in the whole magnetic field. HP-HTS processed sample $x = 0.02$_HIP has almost the same value and similar behaviour as that of the sample processed by CSP ($x = 0.02$). The sample $x = 0.01$_HIP has a slightly enhanced $J_c$ value in the magnetic field range up to 1 T, whereas the $J_c$ value in the high field range is reduced very rapidly and is lower than that of the parent sample. Interestingly, HP-HTS and CSP have the same effect on the $J_c$ value for the low Pr-doping level. We have also included the $J_c$ value for the sample $x = 0.05$ in Figure 9(c), where the $J_c$ value and its magnetic field dependence of this sample are almost the same as those of the sample $x = 0.01$, 0.02. A large amount of Pr-doping, i.e., the sample $x = 0.1$, has started to reduce the $J_c$ value in the whole magnetic field range; however, its behaviour is the same as that of other Pr-doped samples. This analysis suggests that Pr-doping improves the $J_c$ value by around one order of magnitude, especially for the low amount of Pr-doping up to 0.05 prepared by the CSP or HP-HTS processes. However, higher doping reduces the $J_c$ value, which could be due to the presence of impurity phases as observed for other FBS [43].

To summarize our findings, we have plotted the onset $T_c$, transition width ($\Delta T$), the room temperature resistivity ($\rho_{300K}$), $RRR$, and the critical current density ($J_c$) for samples with various Pr-doping contents. In the resistivity measurements, there exists an onset transition temperature ($T_c^{onset}$) at which superconducting transitions commence, and an offset transition temperature ($T_c^{offset}$) at which the resistivity reaches a zero-resistivity value. A difference between $T_c^{onset}$ and $T_c^{offset}$ yields the transition width ($\Delta T$), consistent with definitions in other reports [17], [51]. Here, the $T_c^{onset}$ is determined at the temperature corresponding to 90% of the normal state resistivity, whereas the $T_c^{offset}$ is identified at the temperature corresponding to 10% of the normal state resistivity. For these samples, the normal state resistivity is considered at a temperature near the superconducting transition, specifically at 20 K. Figure 10 depicts the samples processed by CSP, and Figure 11 is for HP-HTS. In these graphs, the data for Gd-added $FeSe_{0.5}Te_{0.5}$ are also included to compare the effect of Gd, which is slightly smaller ion than Fe one. The variation of the onset $T_c$ is depicted in Figure 10(a). There is around 2.5 K suppression of $T_c$ just for a very small amount of Pr-doping ($x$ = 0.01), while the $T_c$ is almost constant around 14.1 K for the doping contents from 0.02 to 0.1. Further higher Pr-doping, i.e., $x$ = 0.2 and 0.3, has a 2 K lower the onset $T_c$ than that of these samples, but zero resistivity has not reached there. These results are the same as for Gd-added $FeSe_{0.5}Te_{0.5}$ samples, as shown in Figure 10(a). Figure 10(b) depicts the transition width variation with doping contents. The parent compound has a transition width of 3 K. Interestingly, Pr-doped samples have reduced the transition width. However, zero resistivity is not observed for higher Pr-doped samples, i.e., $x$>0.1, due to the presence of a huge amount of the impurity. The improvement of the sharpness transition is also observed with Gd-addition, and its behaviour is the same as that of Pr-doped $FeSe_{0.5}Te_{0.5}$. The room temperature resistivity ($\rho_{300K}$) variation with doping contents is shown in Figure 10(c). The $\rho_{300K}$ value is almost the same for all Pr-doping up to 0.1 and for the parent sample. This resistivity value enhances very rapidly for $x$ = 0.2 and 0.3, which again suggests the presence of a large amount of impurity. Interestingly, Gd-addition also follows the same trend of $\rho_{300K}$ as that for Pr -doping contents. The variation of $RRR$ is shown in Figure 10(d), which tells about the homogeneity of the samples. This graph clearly depicts that $RRR$ is reduced with Pr-doping or Gd-addition at Fe sites and suggests that the inhomogeneity of the samples is increased with Pr-doping and Gd-additions. It is well agreed with the above discussion on the microstructural and structural analysis. The calculated $J_c$ at 0 T and 3 T are depicted in Figure 10(e). Low Pr-doped samples have improved $J_c$ value in the whole magnetic field range, whereas high amounts of Pr-doping i.e., $x \geq 0.07$ cause a decrease in the $J_c$ value in the whole magnetic field range due to reduced sample quality. The improvement of $J_c$ behaviour is also reported for Gd-additions, as shown in Figure 10(e).

The high-pressure growth effects for these samples are summarized in Figure 11. The onset $T_c$ of the parent sample is enhanced up to 16.3 K by HP-HTS, which is around 1.5 K higher than that of the parent sample prepared by CSP, as depicted in Figure 11(a). However, Pr-doped samples prepared by HP-HTS have almost the same onset value as the sample prepared by CSP. Gd-added $FeSe_{0.5}Te_{0.5}$ also follows the same behaviour as Pr-doped samples. The transition width of the parent sample is around 3.2 K, which is reduced for the sample $x$ = 0.02_HIP as Gd-added Fe(Se, Te), but unfortunately, other samples such as $x$ = 0.01_HIP, 0.05_HIP, and 0.1_HIP have not depicted the zero resistivity, i.e., the offset transition temperature. The room-temperature resistivity is depicted in Figure 11(c), which is increased with Pr-doping due to the increased impurity phases. The Gd-added sample has also enhanced the resistivity, but it is slightly lower than that of Pr-doped samples. *RRR* values are reduced with Pr-doping contents as depicted in Figure 11(d). This behaviour is the same as observed for the samples grown by CSP. Figure 11(e) depicts the $J_c$ at 0 T and 3 T with doping contents where Gd-additions data are also included. Interestingly, the HP-HTS process of the parent sample has enhanced the $J_c$ by two orders of magnitude compared to CSP, but Pr-doped samples processed with HP-HTS have almost the same $J_c$ value as that of CSP. Normally, the HP-HTS process enhances the $J_c$ value due to improved grain connections and sample density. But the Pr-doped samples have almost the same $J_c$ value as the CSP process. It could be possible because HP-HTS enhances the sample density, but at the same time, the impurity phase is also increased. That's why, $J_c$ is almost constant for a low Pr-doped sample but the impurity phase is enhanced very rapidly by HP-HTS for a high amount of Pr-doping, i.e., $x > 0.07$, and because of this, the effect of HP-HTS cannot be seen in these samples. This analysis is well in agreement with the results of the Gd-added samples [17]. Our analysis suggests that rare earth doping or additions in $FeSe_{0.5}Te_{0.5}$ are not affecting too much on the $T_c$ for bulks either processed by CSP or HP-HTS. Interestingly, the $J_c$ value is improved by around one order of magnitude by this rare earth addition and most effective doping or addition is up to 10%. These results are well in agreement with the reported rare earth addition in $MgB_2$ samples [13].

## Conclusion

Pr-doping at Fe sites in $FeSe_{0.5}Te_{0.5}$ bulks has been studied by the CSP and HP-HTS processes, and different characterizations have been performed. Structural analysis of the $Pr_{1-x}Fe_xSe_{0.5}Te_{0.5}$ bulks processed either through CSP or HP-HTS confirmed a small increment in the lattice parameters '*a*' and '*c*', and the reduced or almost the same impurity phase has been observed up to 10% Pr-doping level compared to the parent $FeSe_{0.5}Te_{0.5}$ ($x = 0$), which confirms the successful insertion of the rare earth in the tetragonal lattice of Fe(Se, Te). Furthermore, the substitution of Pr atoms in the superconducting lattice's interlayer is suggested by Raman spectroscopy results and DFT calculations. Remarkably, the elemental mapping analysis suggests that Pr-doping enhances the inhomogeneity of the constituent elements, but does not affect too much the onset transition temperature; even higher Pr-doping levels ($x$>0.1) have a lower the onset $T_c$ of 2 K than that of the parent sample, where the hexagonal phase is observed at more than 40%. The critical current density of the parent sample ($x = 0$) has been enhanced by one order of magnitude by Pr-doping in the whole magnetic field range, which is almost the same for the doping range from 0.02 to 0.1. The HP-HTS process enhances $J_c$ of the parent sample ($x$ = 0_HIP) by two orders of magnitude more than that of the sample $x = 0$. However, Pr-doped samples prepared by HP-HTS have reduced the sample density in contrast to the parent sample grown under high-pressure conditions, but almost have the same $J_c$ value as the samples prepared by CSP. A comparative study of these results also well agreed with the Gd-added $FeSe_{0.5}Te_{0.5}$. Our analysis confirms that the rare earth doping or addition in $FeSe_{0.5}Te_{0.5}$ does not affect much the onset of the superconducting transition temperature of Fe(Se, Te), but it enhances the $J_c$ value in the whole magnetic field irrespective of the growth methods, and this effect is more observable up to the 10% rare earth doping. This information can be very useful for the practical applications of these superconducting materials, and it is also demanding more research in this direction, especially for very low amounts of rare earth doping or addition ($x \leq 0.10$).

### CRediT authorship contribution statement

**Priya Singh:** Writing– review & editing, Investigation, Formal analysis. **Manasa Manasa:** Writing – review & editing, Writing – original draft, Investigation, Formal analysis, Data curation. **Mohammad Azam:** Writing – review & editing, Data curation. **Tatiana Zajarniuk:** Data curation. **Svitlana Stelmakh:** Formal analysis, Data curation. **Taras Palasyuk:** Writing

– review & editing, Formal analysis, Data curation. **Jan Mizeracki:** Data curation. **Tomasz Cetner:** Writing – review & editing, Data curation. **Andrzej Morawski:** Data curation. **Cezariusz Jastrzębski:** Writing – review & editing, Data curation. **Michał Wierzbicki:** Formal analysis, Writing – review & editing. **Shiv J. Singh:** Writing – review & editing, Writing – original draft, Visualization, Validation, Supervision, Software, Resources, Methodology, Investigation, Funding acquisition, Formal analysis, Conceptualization.

## Declaration of competing interest

The authors declare that they have no known competing financial interests or personal relationships that could have appeared to influence the work reported in this paper.

## Data availability

The raw/processed data required to reproduce these findings cannot be shared at this time due to technical or time limitations. Data are available upon request to the corresponding author.

## Acknowledgments:

The work was funded by SONATA-BIS 11 project (Registration number: 2021/42/E/ST5/00262) sponsored by National Science Centre (NCN), Poland. SJS acknowledges financial support from National Science Centre (NCN), Poland through research Project number: 2021/42/E/ST5/00262.

**Table 1:**

List of the sample codes and quantitative analysis of impurity phases of $Pr_xFe_{1-x}Se_{0.5}Te_{0.5}$ bulks prepared by CSP and HP-HTS (denoted HIP) processes.

| **Sample Code** | **Hexagonal Phase (%)** | **PrSe Phase (%)** |
|---|---|---|
| $x = 0$ | ~5 | _ |
| $x = 0.01$ | ~1 | _ |
| $x = 0.02$ | ~3 | _ |
| $x = 0.03$ | ~3 | _ |
| $x = 0.05$ | ~3 | _ |
| $x = 0.07$ | ~3 | _ |
| $x = 0.1$ | ~7 | ~2 |
| $x = 0.2$ | ~35 | ~4 |
| $x = 0.3$ | ~45 | ~13 |
| $x = 0$_HIP | ~ 6 | _ |
| $x = 0.01$_HIP | ~3 | _ |
| $x = 0.02$_HIP | ~4 | _ |
| $x = 0.05$_HIP | ~2 | _ |
| $x = 0.1$_HIP | ~12 | _ |

**Table 2:** List of the molar ratio of various constituent elements for $Pr_xFe_{1-x}Se_{0.5}Te_{0.5}$ prepared by CSP and HP-HTS (denoted as HIP) processes.

| Sample Code | Fe Molar Ratio | Se Molar ratio | Te Molar Ratio | Pr Molar Ratio |
|---|---|---|---|---|
| $x$ = 0 | 1 | 0.49 | 0.5 | - |
| $x$ = 0.02 | 1 | 0.49 | 0.49 | 0.019 |
| $x$ = 0.05 | 1 | 0.49 | 0.5 | 0.045 |
| $x$ = 0.1 | 1 | 0.63 | 0.6 | 0.11 |
| $x$ = 0.2 | 1 | 0.5 | 0.53 | 0.17 |
| $x$ = 0.3 | 1 | 0.55 | 0.69 | 0.22 |
| $x$ = 0_HIP | 1 | 0.5 | 0.51 | - |
| $x$ = 0.02_HIP | 1 | 0.49 | 0.48 | 0.02 |
| $x$ = 0.05_HIP | 1 | 0.49 | 0.51 | 0.059 |
| $x$ = 0.1_HIP | 1 | 0.5 | 0.51 | 0.110 |

**Figure 1:** Powder X-ray diffraction (XRD) patterns of $Pr_xFe_{1-x}Se_{0.5}Te_{0.5}$ bulks ($x$ = 0, 0.01, 0.02, 0.03, 0.05, 0.07, 0.1 and 0.2 and 0.3) prepared at the room temperature by **(a)** CSP and **(b)** HP-HTS process. The variation of the calculated lattice parameters **(c)** '*a*' **(d)** '*c*' and **(e)** the lattice volume ($V$) with the nominal Pr substitution level ($x$) for all Pr-doped samples either prepared by CSP (**closed symbol**) or HP-HTS (**open symbol**) process. The lattice parameters and lattice volume for Gd-added $FeSe_{0.5}Te_{0.5}$ are also included in the previously published paper [17] in Figure 1(c), (d) and (e). The tetragonal phase of $FeSe_{0.5}Te_{0.5}$ was observed as the superconducting phase. The hexagonal phase, $Fe_7Se_8$, has been found and is depicted as "H" in figures (a) and (b). Table 1 contains a list of the samples and the obtained impurity phases observed from this XRD analysis.

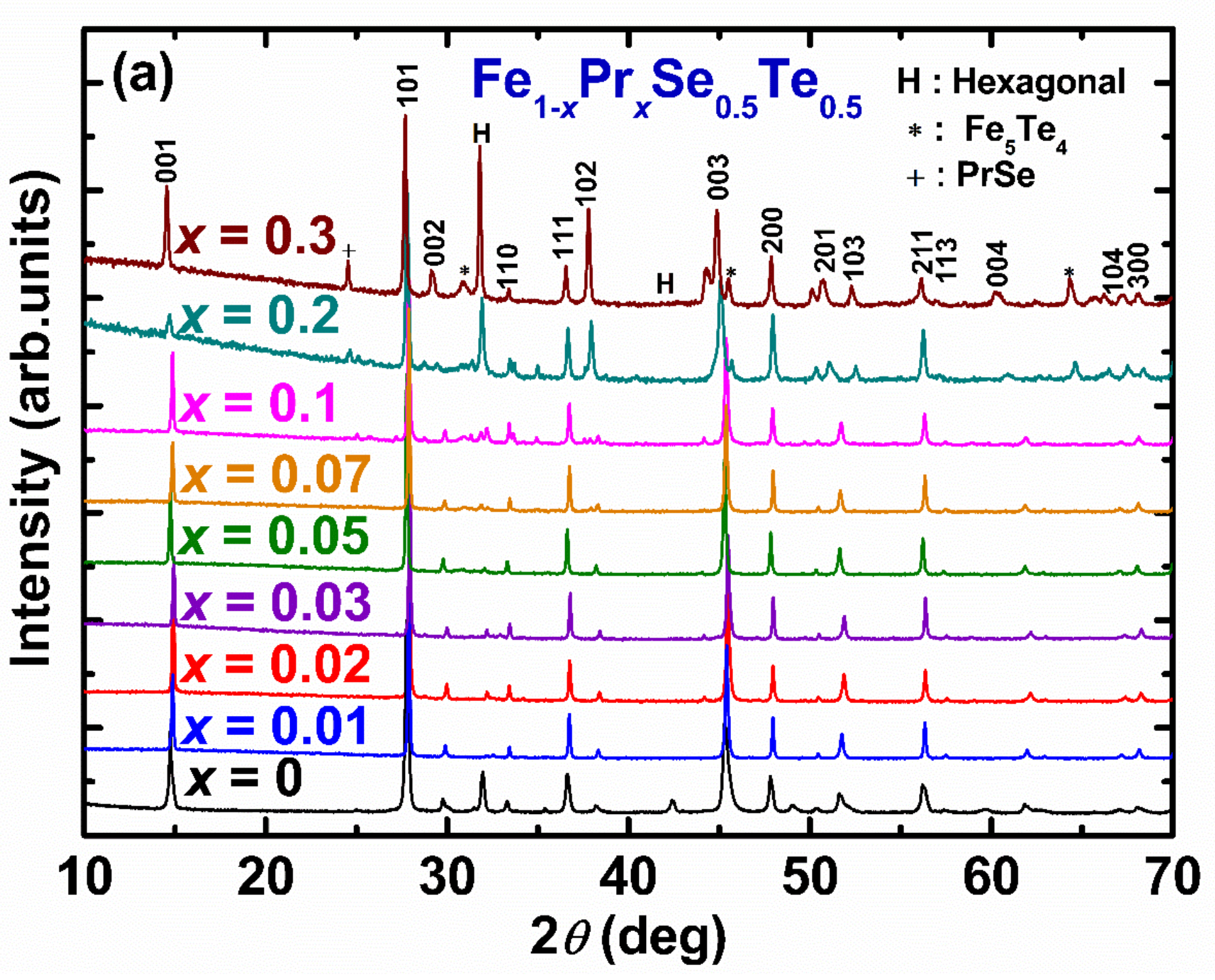

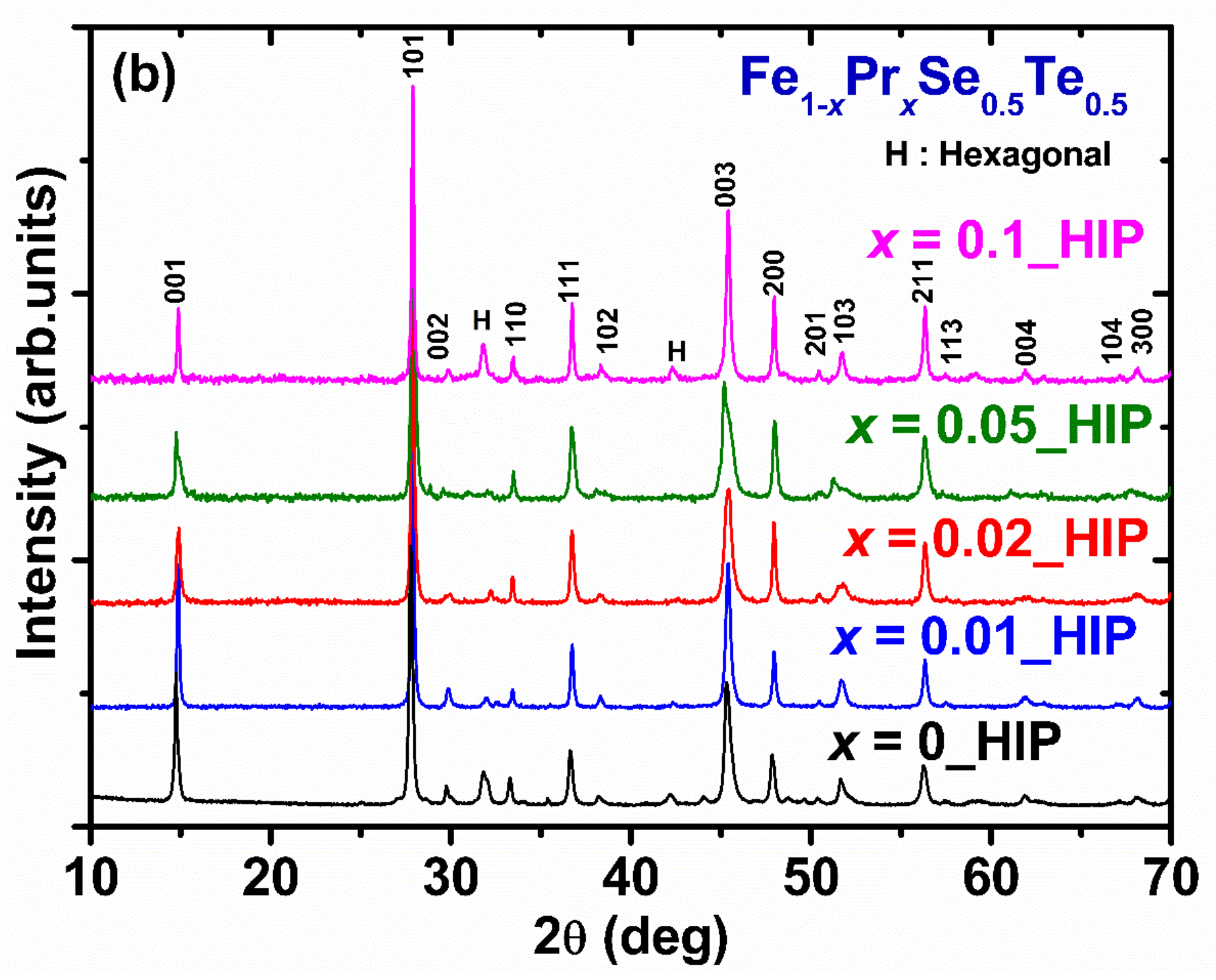

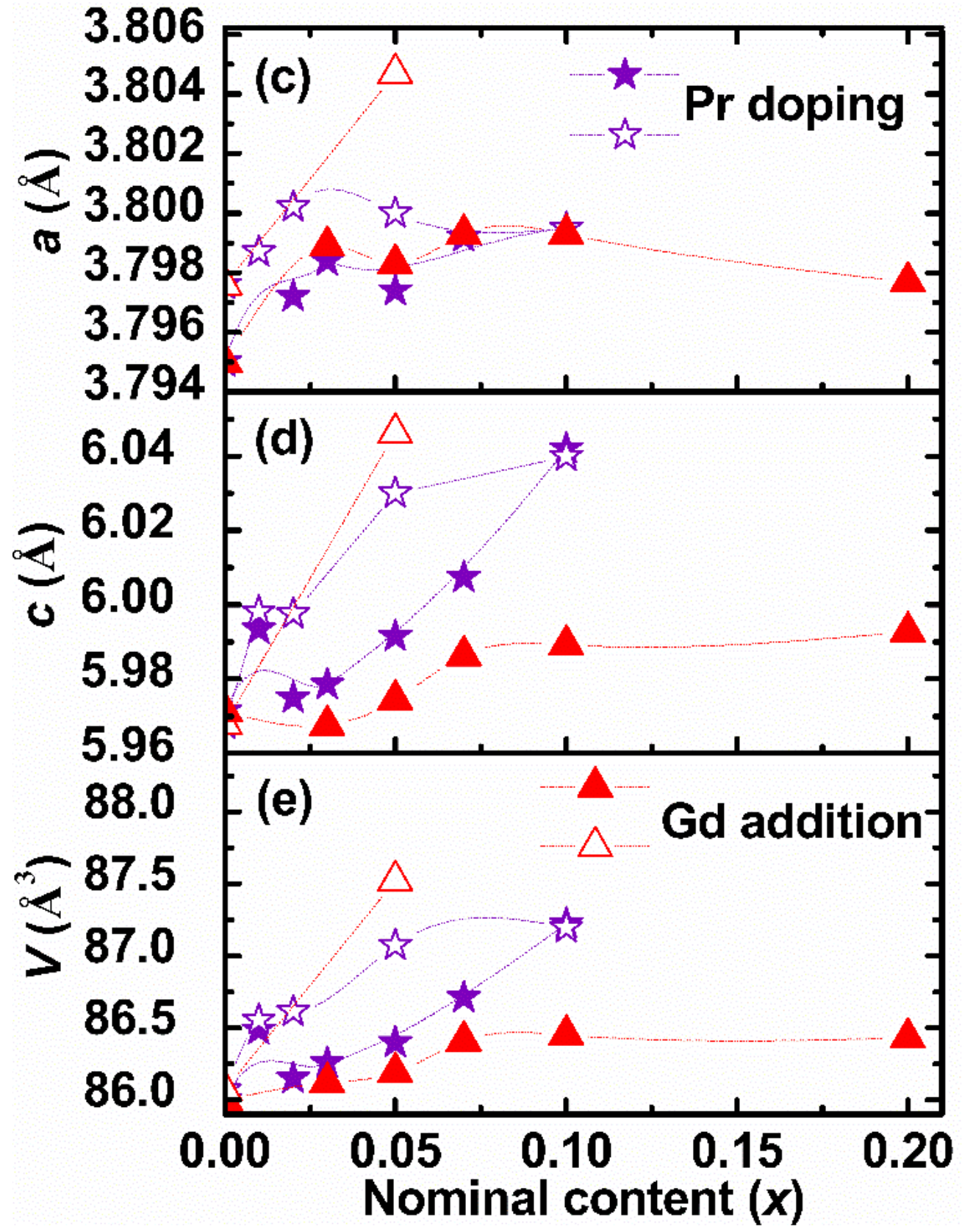

**Figure 2:** Raman scattering study of polycrystalline $FeSe_{0.5}Te_{0.5}$ doped with Pr. **(a)** Experimental Raman spectrum collected at room temperature from the surface of parent $FeSe_{0.5}Te_{0.5}$ using high laser power (4 mW). Green spectrum from the paper of Zargar et al 2015 [30] illustrates one of many examples of misinterpreted phonon assignments for $FeSe_{0.5}Te_{0.5}$. **(b)** Experimental Raman spectrum collected at low-temperature, under gaseous nitrogen atmosphere using medium laser power (0.8 mW). **(c)** Representative low-temperature Raman spectra of $FeSe_{0.5}Te_{0.5}$ with addition of Pr 5 at. % ($x$ = 0.02) acquired using low and medium laser power of 0.08 mW and 0.8 mW (red and blue spectrum respectively). The assignment of detected signals related to molecular rotation of nitrogen and lattice vibrations of Pr-doped $FeSe_{0.5}Te_{0.5}$ is shown for the spectrum collected at medium laser power. For comparison, positions of Fe related lattice modes in both spectra are indicated by arrow headed lines. An enlarged fragment of the red spectrum along with peaks fitted by Lorentz shape are shown in the inset. **(d)** Raman frequencies of respective lattice modes measured in parent compound ($FeSe_{0.5}Te_{0.5}$) and materials with Pr addition (5 and 10 at. %) were observed for two series of samples synthesized at ambient and high-pressure conditions (full and empty symbols respectively). Experimental data on the frequency of Se/Te and Fe related phonons modes from earlier reported studies for single crystalline materials $Fe_{0.95}Se_{0.56}Te_{0.44}$ [31], $Fe_{1.09}Te$ [31] and $FeSe_{0.5}Te_{0.5}$ [32] are included for comparison.

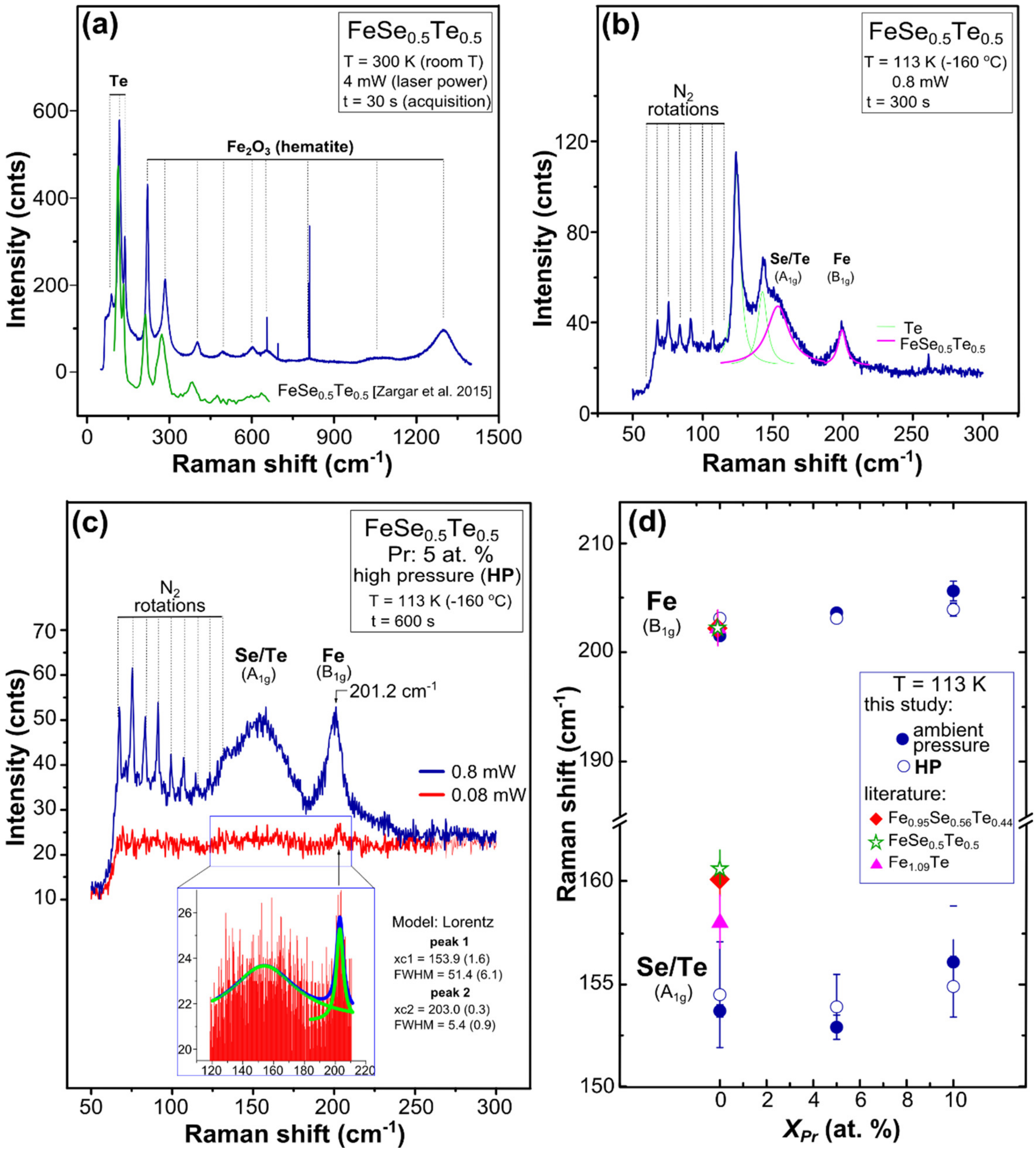

**Figure 3:** Elemental Mapping of the constituent elements of various $Pr_xFe_{1-x}Se_{0.5}Te_{0.5}$ bulks prepared by CSP: **(i)** Parent $x = 0$ **(ii)** $x = 0.02$ **(iii)** $x = 0.05$ **(iv)** $x = 0.1$ **(v)** $x = 0.3$ samples.

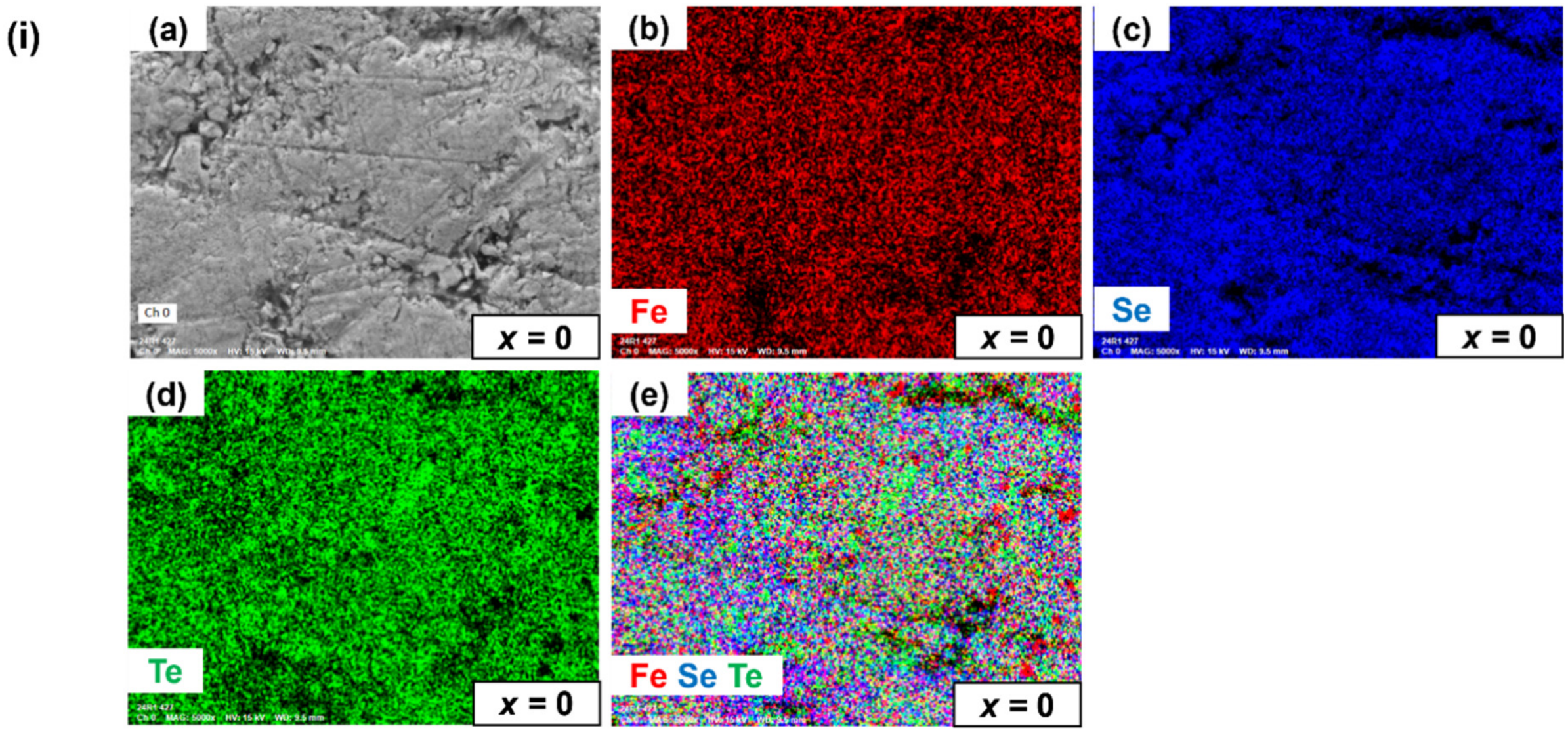

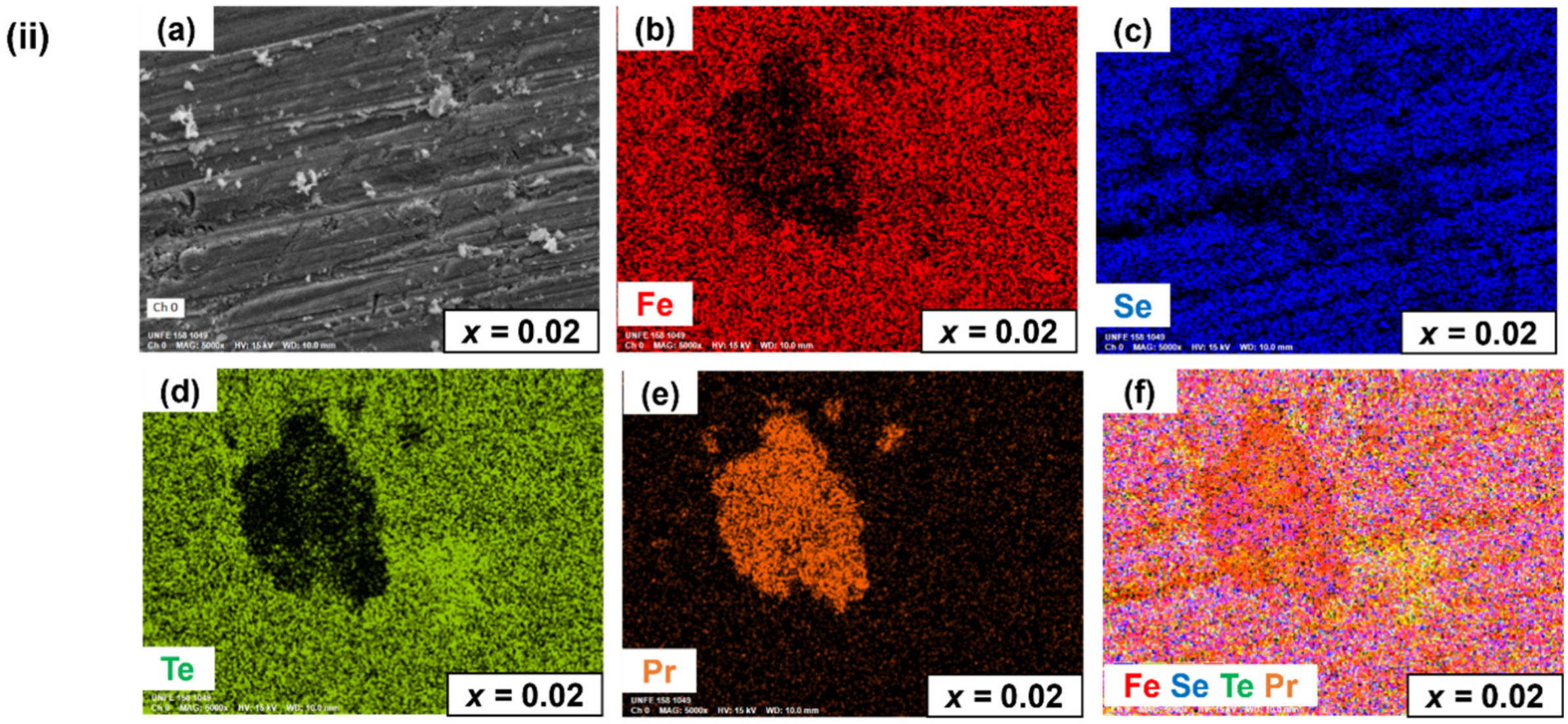

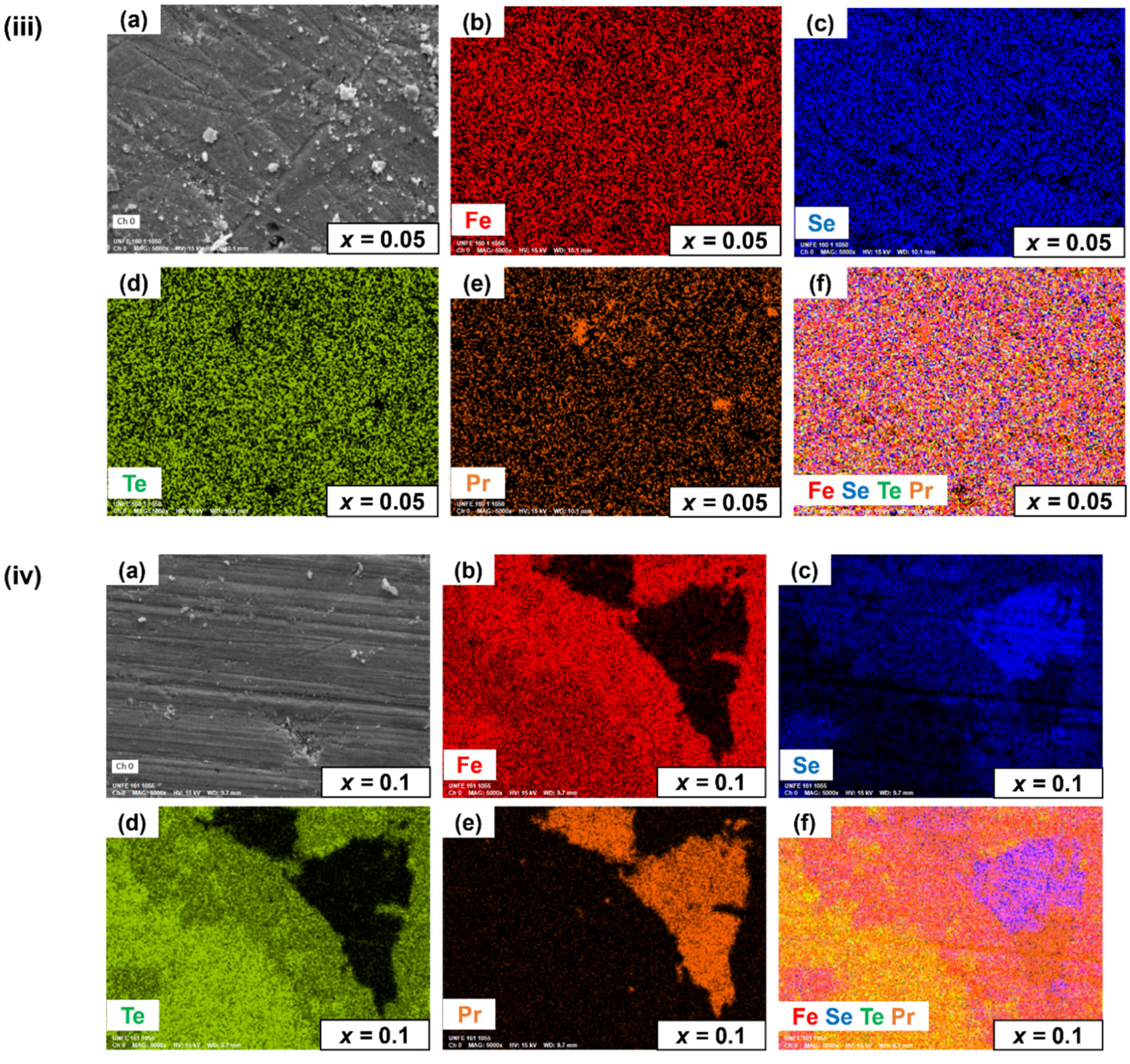

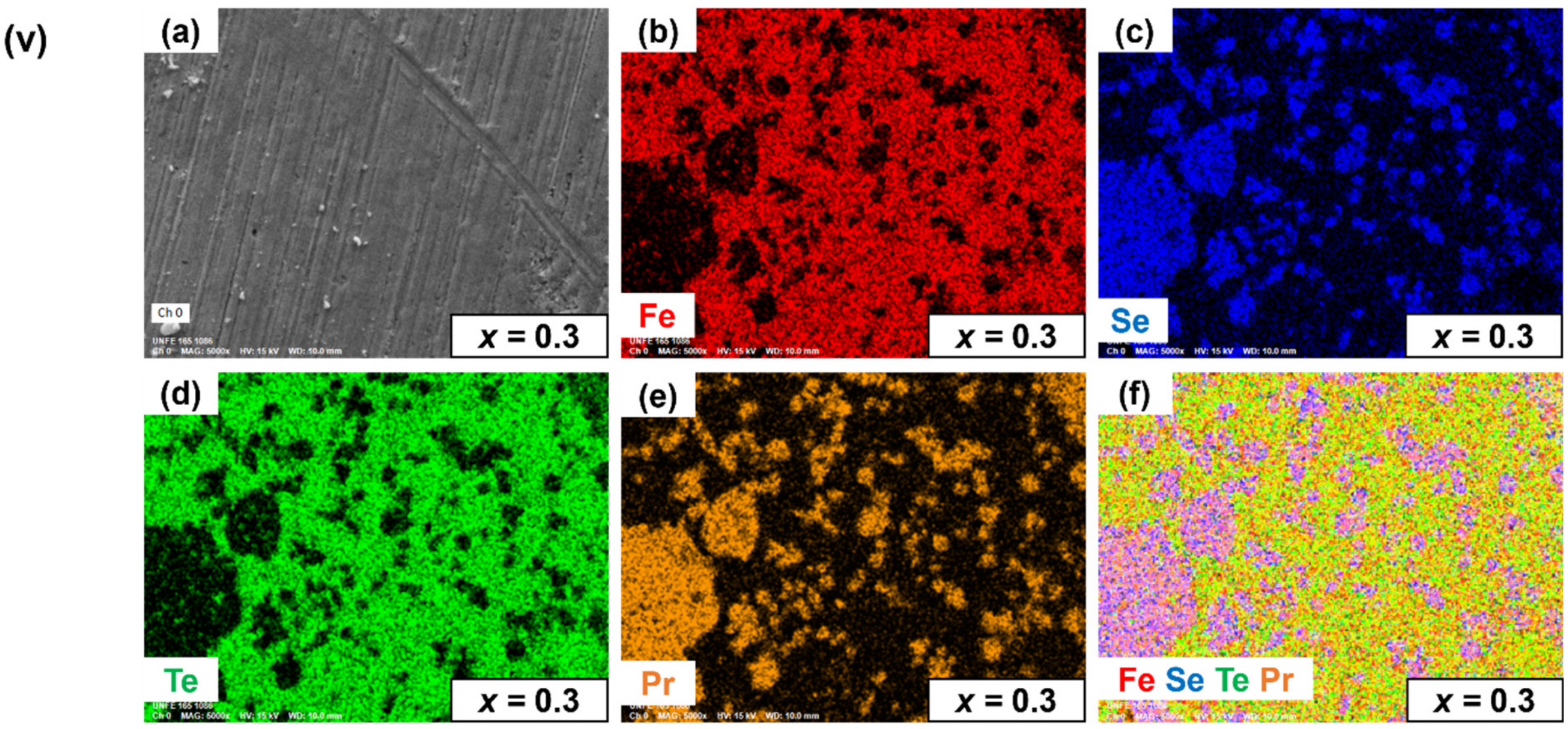

**Figure 4:** Elemental Mapping of the constituent elements of various $Pr_xFe_{1-x}Se_{0.5}Te_{0.5}$ bulks prepared by HP-HTS: **(i)** Parent $x$ = 0_HIP **(ii)** $x$ = 0.02_HIP **(iii)** $x$ = 0.05_HIP **(iv)** $x$ = 0.1_HIP.

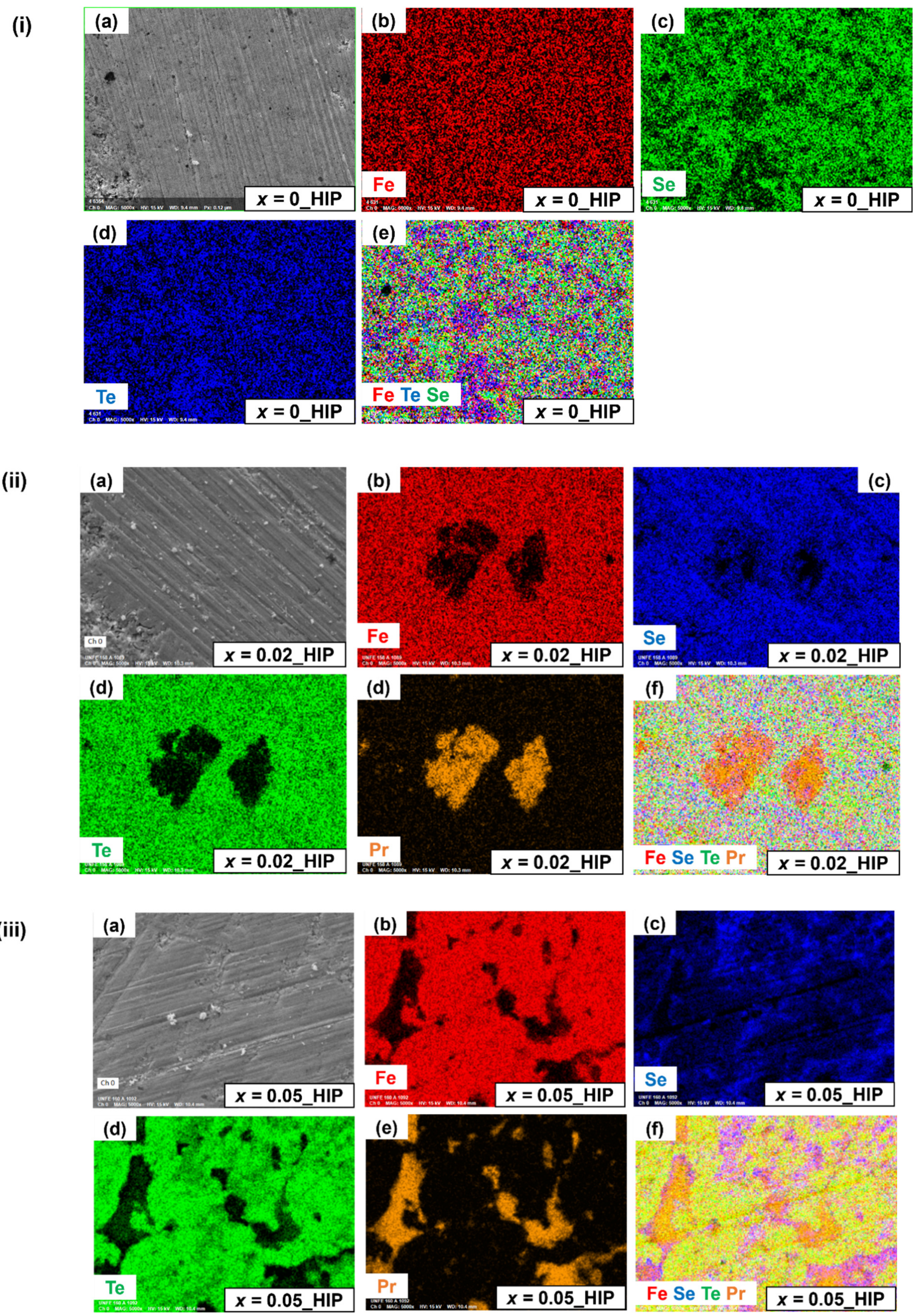

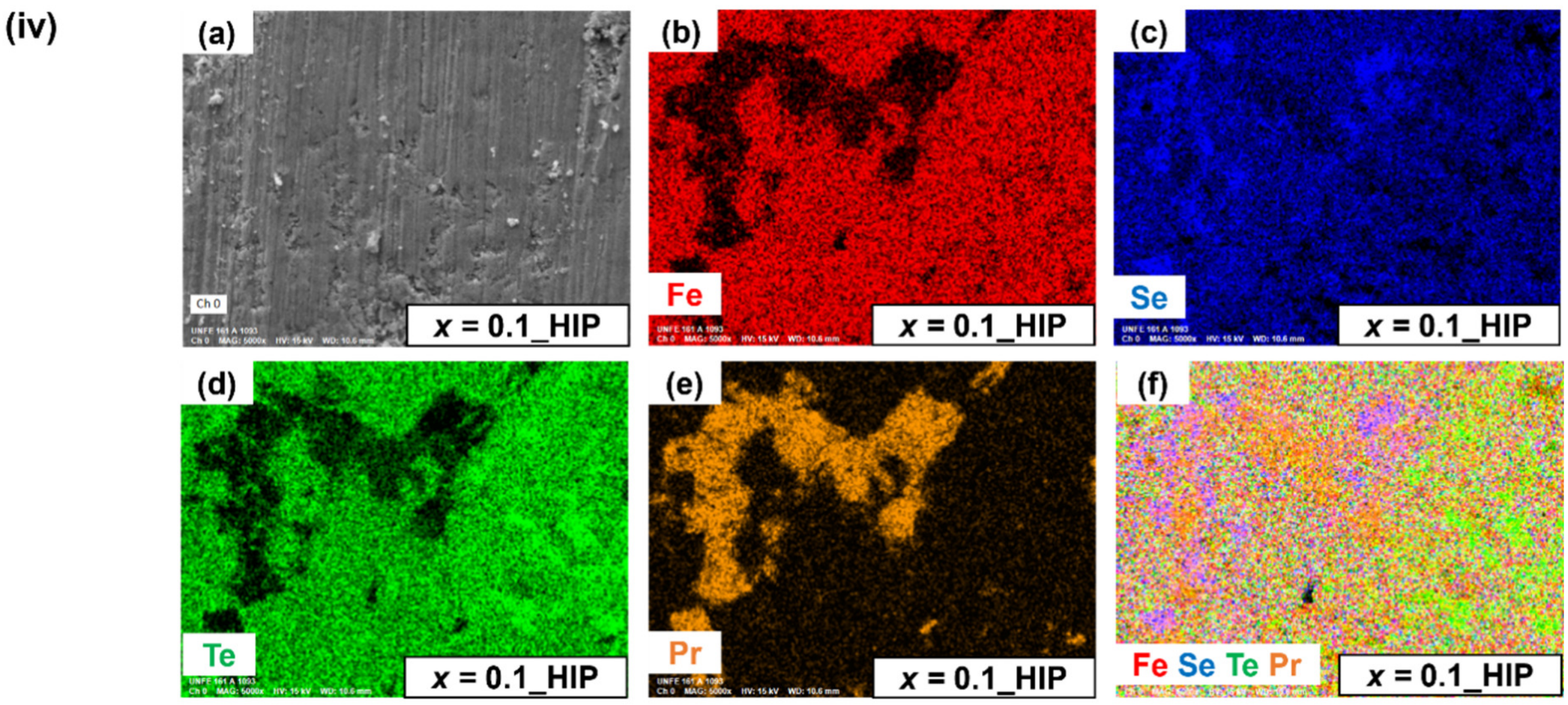

**Figure 5:** Back-scattered (BSE) images of $Pr_xFe_{1-x}Se_{0.5}Te_{0.5}$ polycrystalline samples prepared by CSP where pores and hexagonal phases(H) are marked using arrows: **(a)-(c)** for parent $x = 0$; **(d)-(f)** for $x = 0.02$; **(g)-(i)** for $x = 0.05$; **(j)-(l)** for $x = 0.1$; and **(m)-(o)** $x = 0.3$.

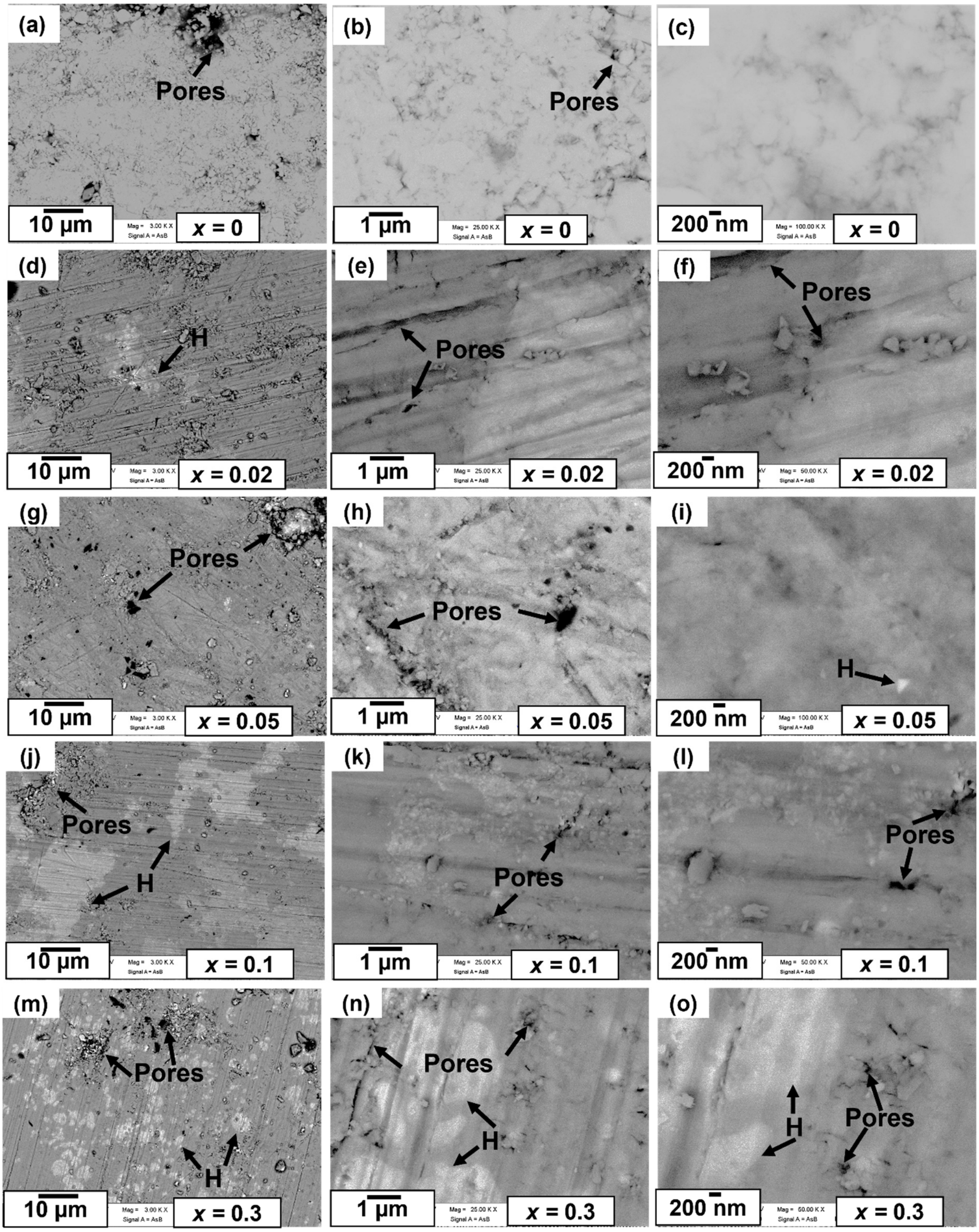

**Figure 6:** Back-scattered (BSE) images of $Pr_xFe_{1-x}Se_{0.5}Te_{0.5}$ polycrystalline samples prepared by HP-HTS where pores and hexagonal phases(H) are marked using arrows: **(a)-(c)** for parent $x$ = 0_HIP; **(d)-(f)** for $x$ = 0.02_HIP; **(g)-(i)** for $x$ = 0.05_HIP; **(j)-(l)** for $x$ = 0.1_HIP.

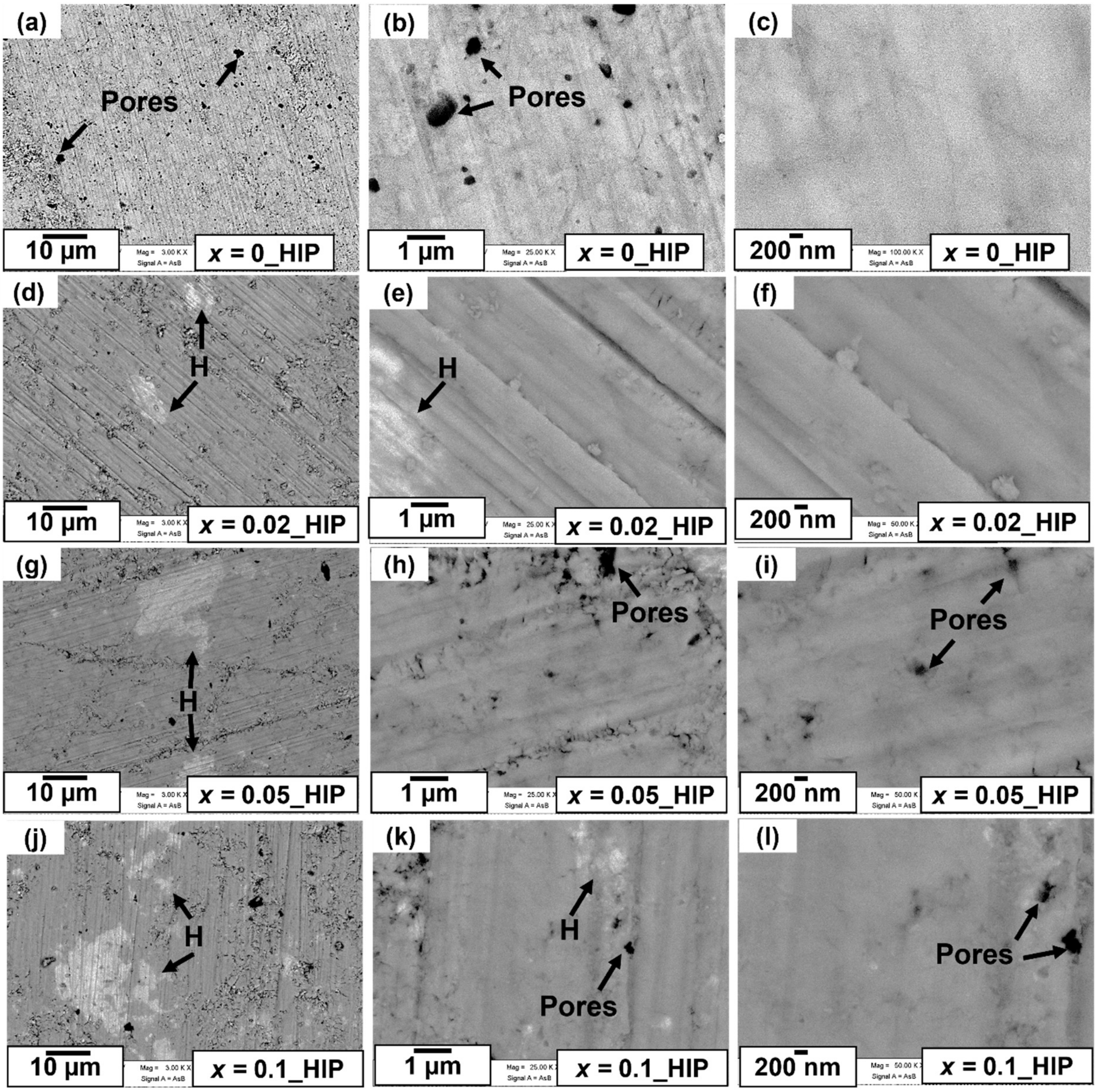

**Figure 7:** **(a)** The temperature dependence of resistivity for $Pr_xFe_{1-x}Se_{0.5}Te_{0.5}$ ($x$ = 0, 0.01, 0.02, 0.03, 0.05, 0.07, 0.1) samples prepared by CSP up to room temperature **(b)** The variation of the resistivity ($\rho$) with the temperature for $Pr_xFe_{1-x}Se_{0.5}Te_{0.5}$ bulks for $x$ = 0.2 and 0.3. The inset figure shows the low-temperature variation of the resistivity for these samples. **(c)** Low-temperature resistivity up to 16 K temperature of $Pr_xFe_{1-x}Se_{0.5}Te_{0.5}$ bulks prepared by CSP **(d)** The temperature variation of the resistivity under different currents ($I$ = 5, 10 and 20 mA) for $Pr_xFe_{1-x}Se_{0.5}Te_{0.5}$ samples $x$ = 0, 0.02, 0.05 and 0.1 prepared by CSP method.

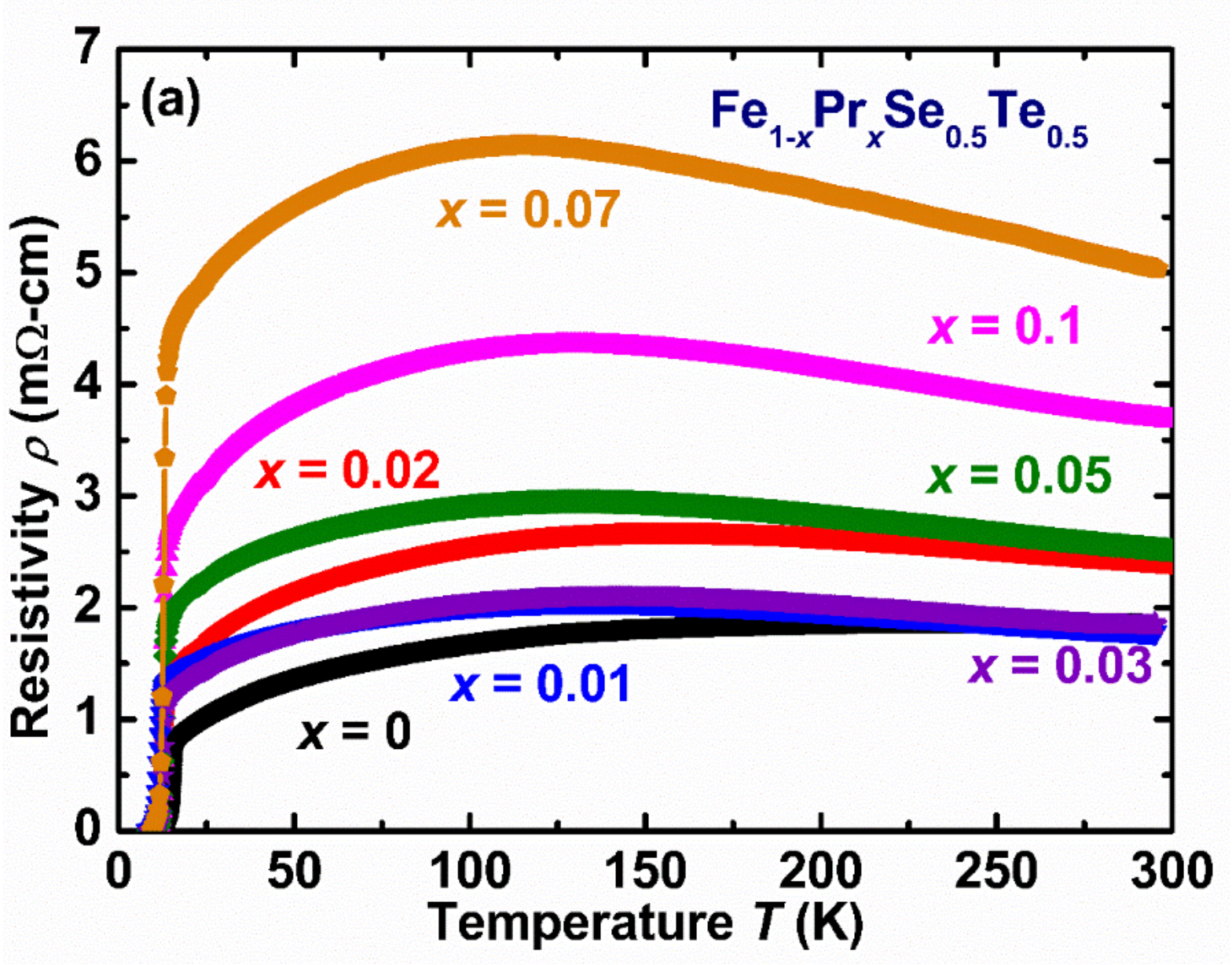

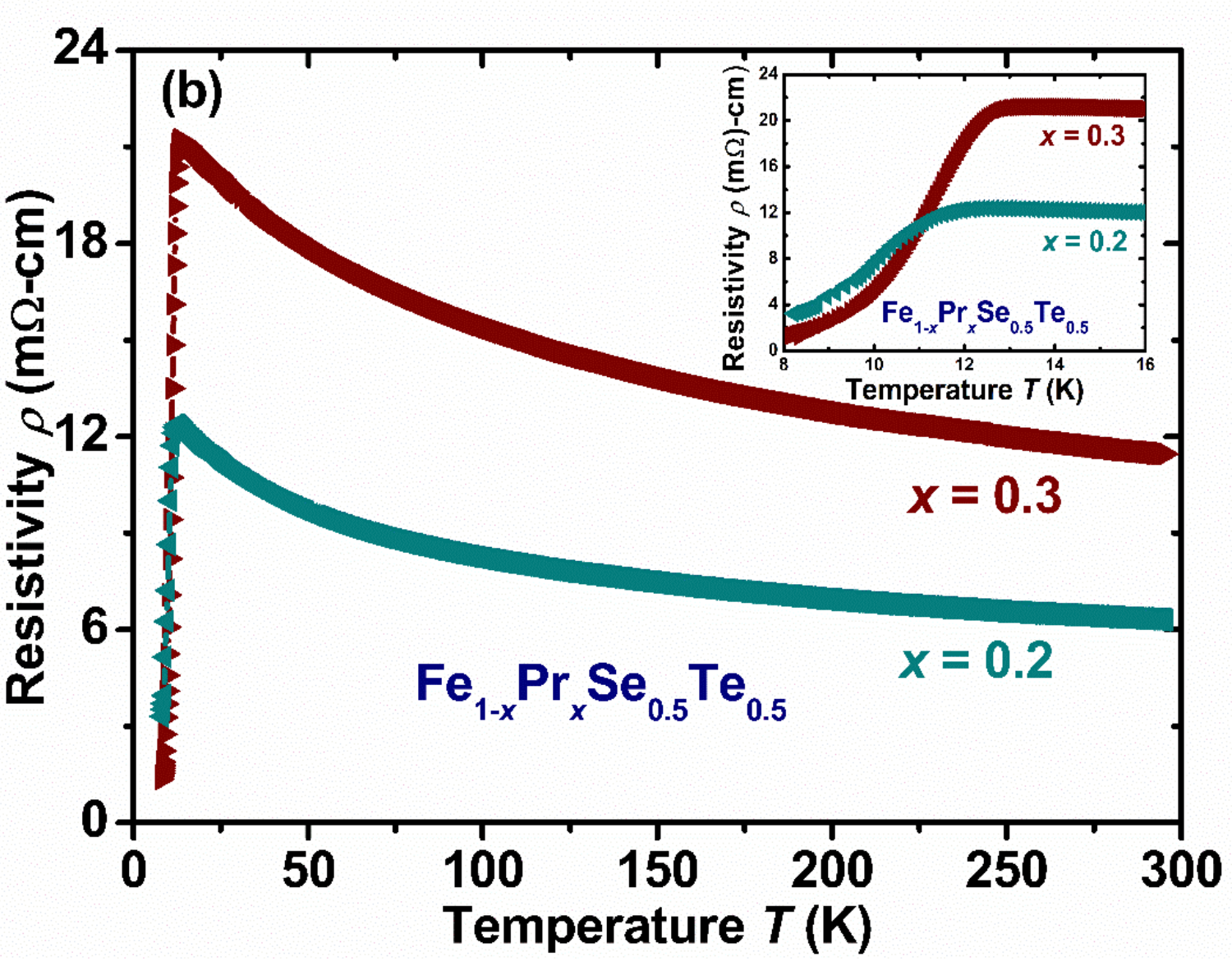

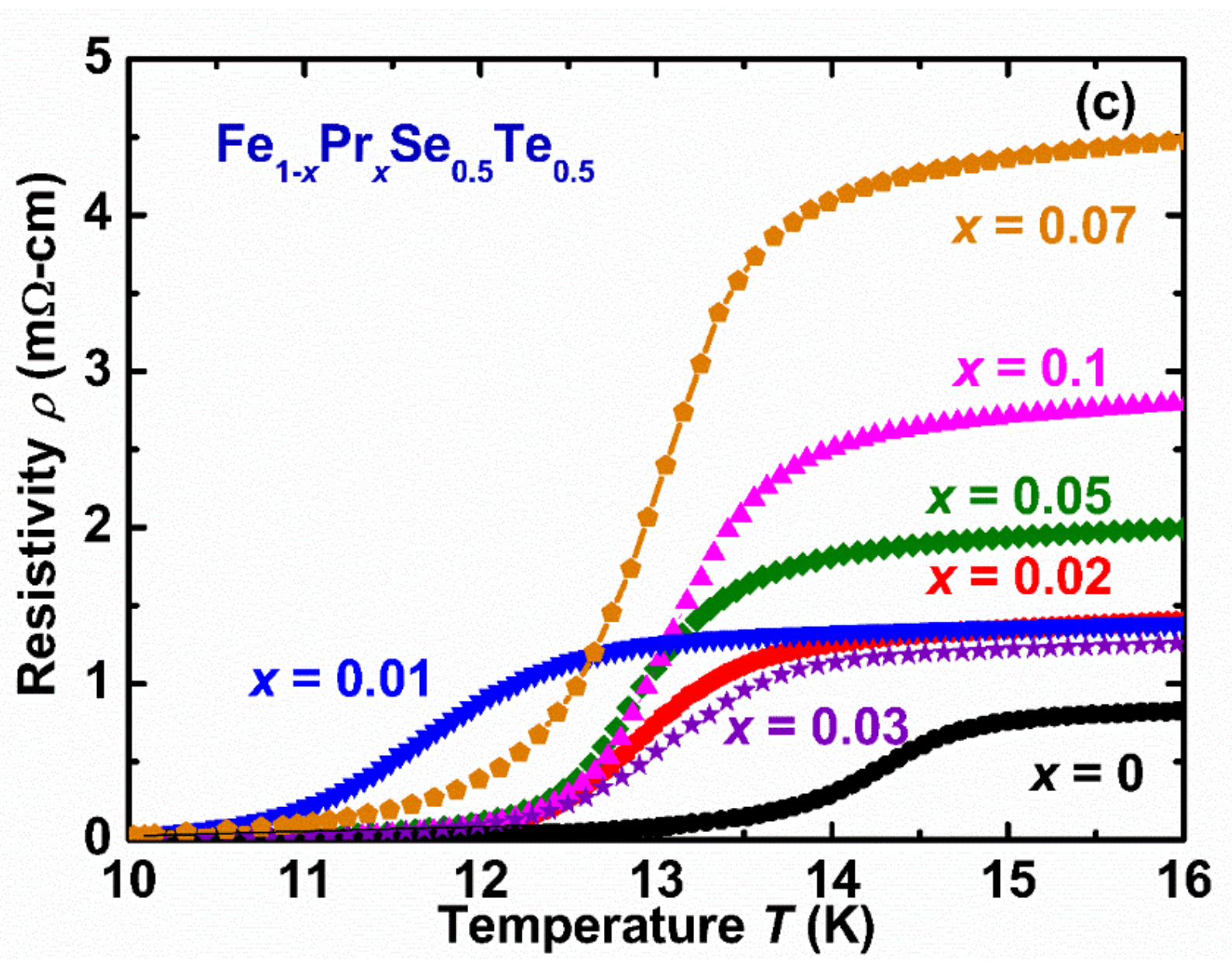

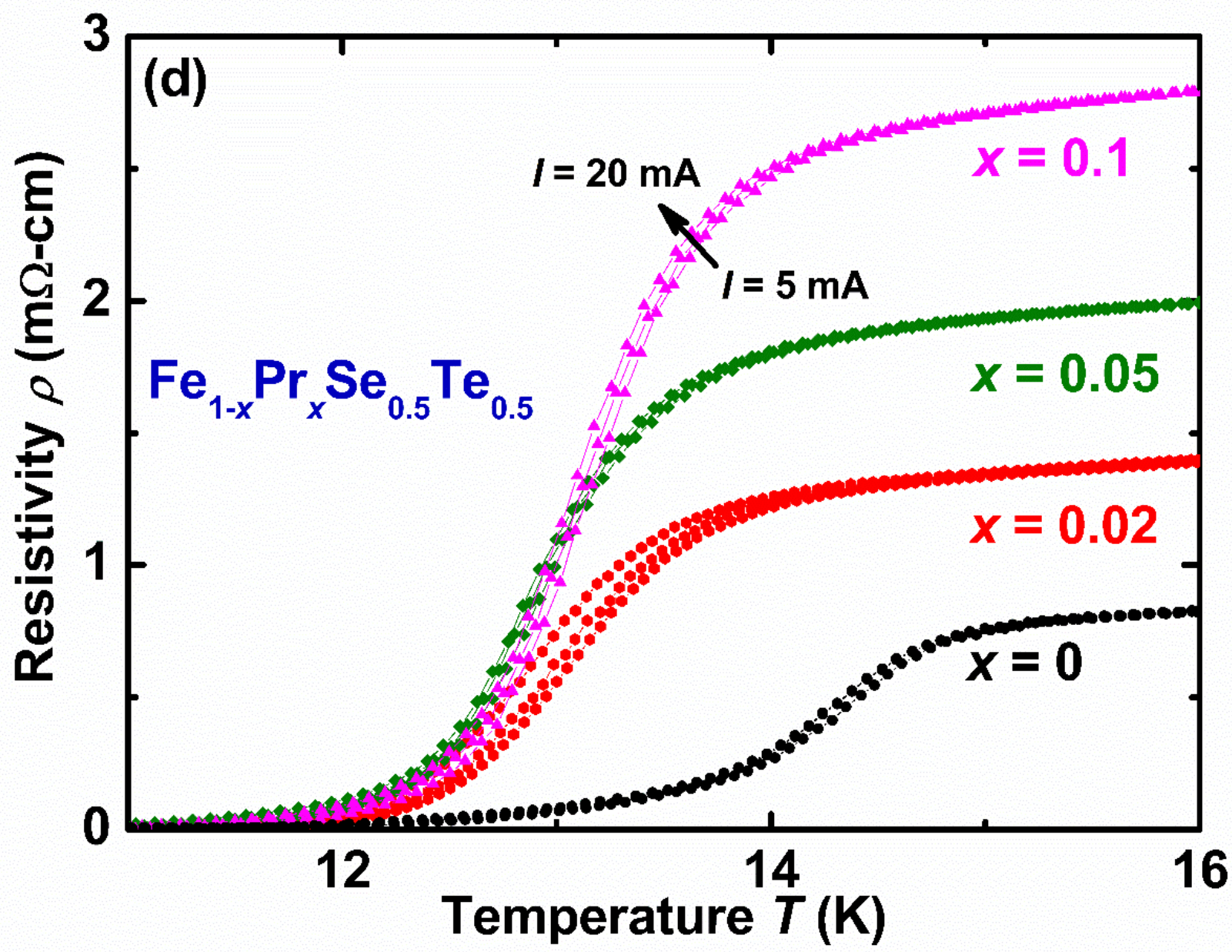

**Figure 8:** (**a**) Samples prepared by HP-HTS: (**a**) the temperature dependence of the resistivity of $Pr_xFe_{1-x}Se_{0.5}Te_{0.5}$ bulks up to room temperature for $x$ = 0, 0.02, 0_HIP and 0.02_HIP. The inset figure shows the low-temperature resistivity of these samples in the temperature range 10-16 K. (**b**) The variation of resistivity of $Pr_xFe_{1-x}Se_{0.5}Te_{0.5}$ bulks for $x$ = 0.01_HIP, 0.05_HIP and 0.1_HIP till 250K (**c**) The temperature dependence of the resistivity under different currents ($I$ = 5, 10 and 20 mA) for $x$ = 0_HIP and 0.02_HIP $x$ = 0_HIP shows no dependence on applied current.

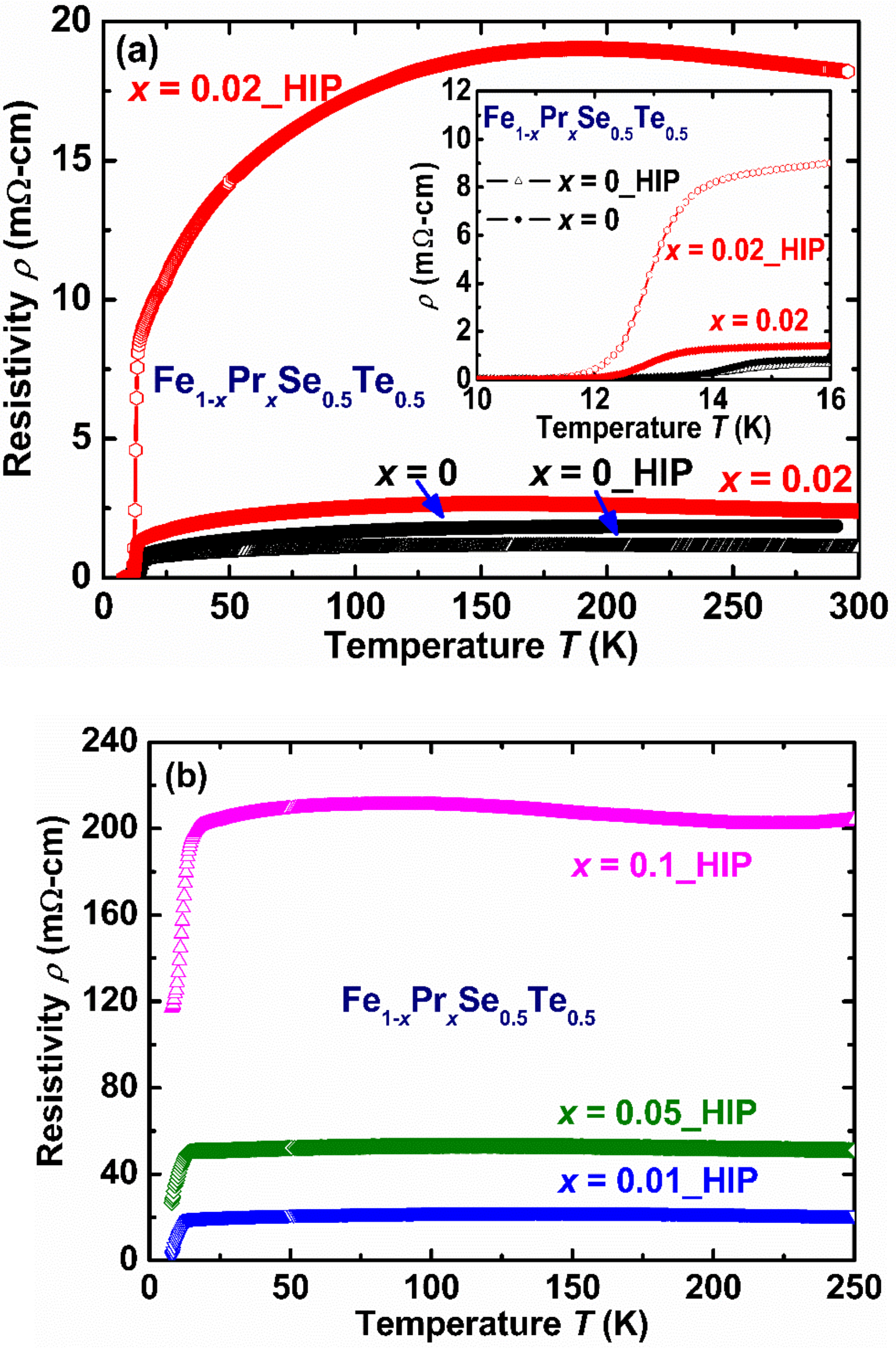

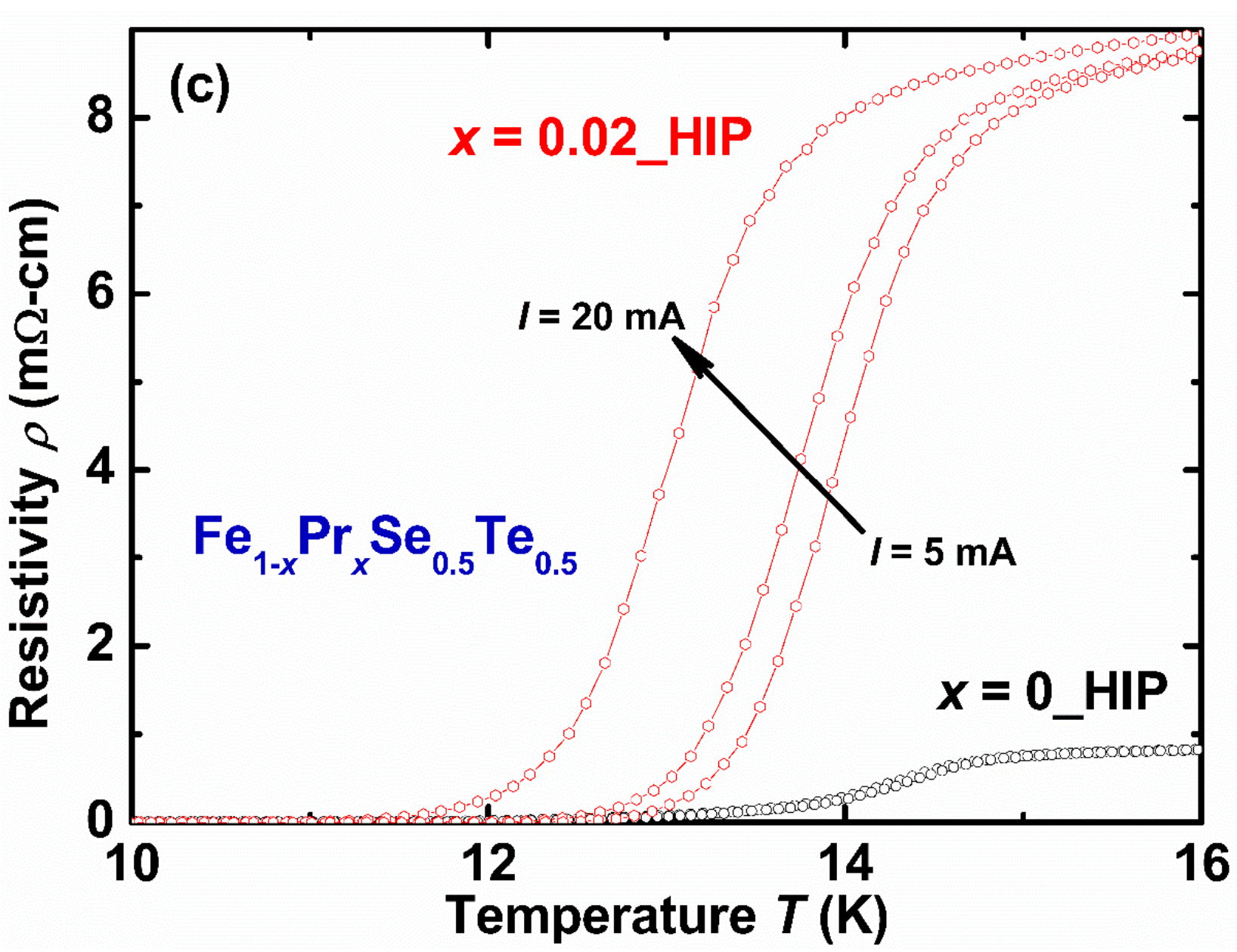

**Figure 9:** The temperature dependence of the normalized magnetic susceptibility ($\chi$ = $4\pi M/H$) measured under zero field-cooled (ZFC) and field-cooled (FC) modes in an applied magnetic field $H$ = 50 Oe for $Pr_xFe_{1-x}Se_{0.5}Te_{0.5}$ bulks prepared by **(a)** CSP and **(b)** HP-HTS. **(c)** The variation of critical current density ($J_c$) with respect to the applied magnetic field for $Pr_xFe_{1-x}Se_{0.5}Te_{0.5}$ bulks ($x$ = 0, 0_HIP, 0.01, 0.01_HIP, 0.02, 0.02_HIP, 0.05 and 0.1) up to 9 T and at 7 K. The inset figure shows the hysteresis loop ($M$-$H$) for the sample $x$ = 0.02 and 0.02_HIP.

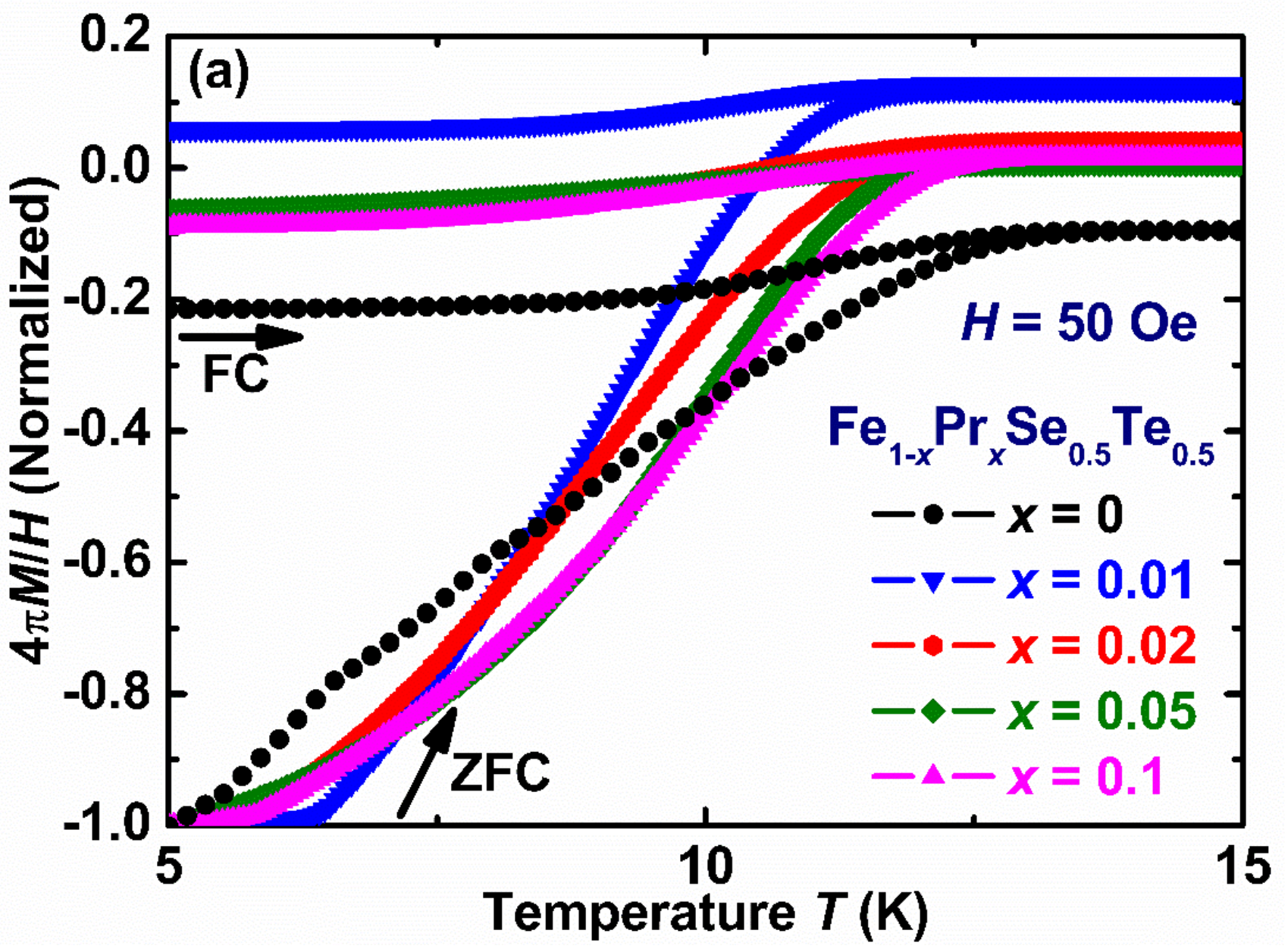

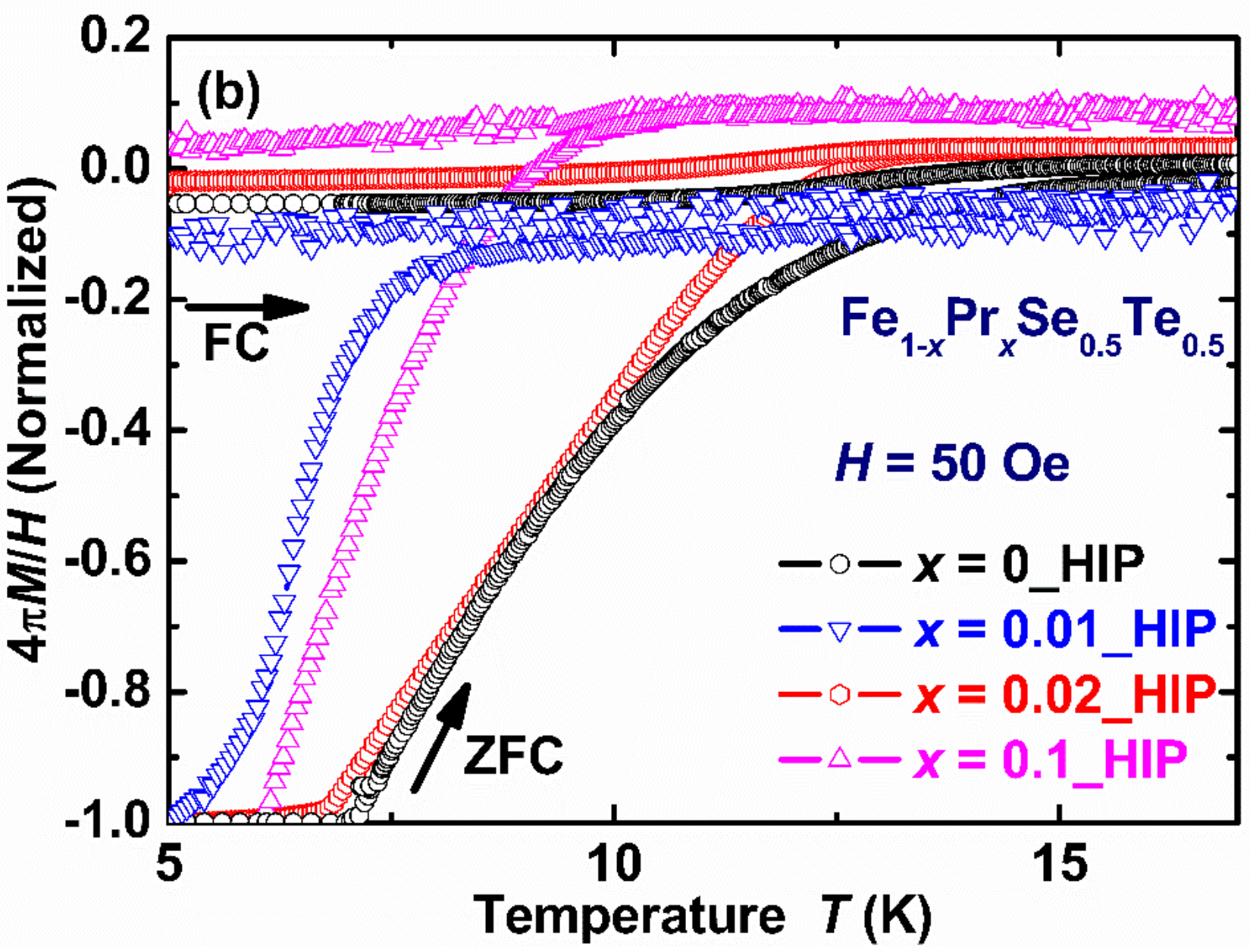

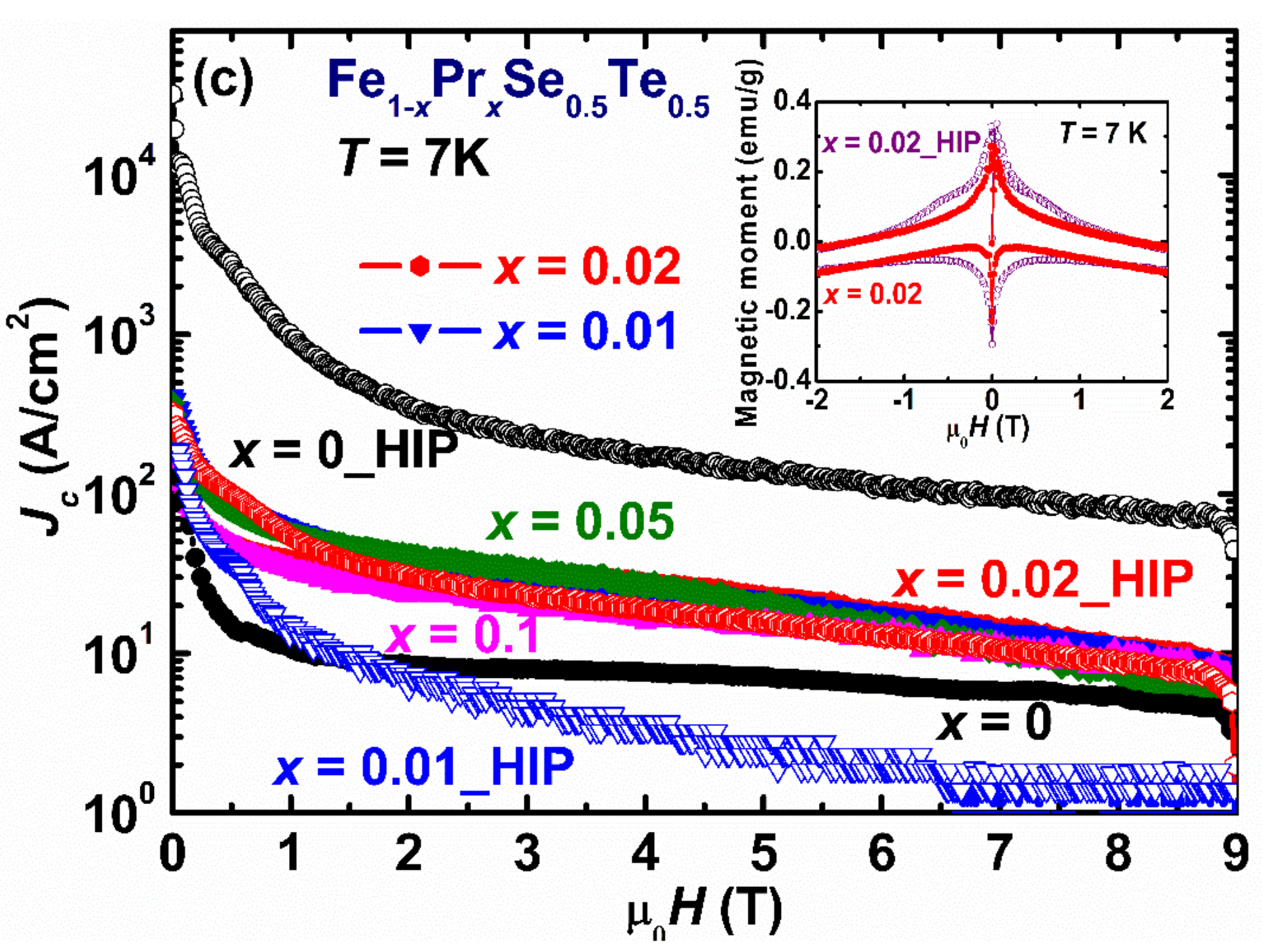

**Figure 10:** The variation of **(a)** the onset transition temperature ($T_c$) **(b)** the transition width ($\Delta T$) **(c)** the room temperature resistivity ($\rho_{300K}$) **(d)** residual resistivity ratio ($RRR = \rho_{300K} / \rho_{20K}$) **(e)** the critical current density ($J_c$) at 7 K for $H$ = 0 T (**closed symbol**) and 3 T (**open symbol**) for $Pr_xFe_{1-x}Se_{0.5}Te_{0.5}$ bulks prepared by CSP with respect to the nominal contents ($x$) of Pr-doping or Gd-additions.

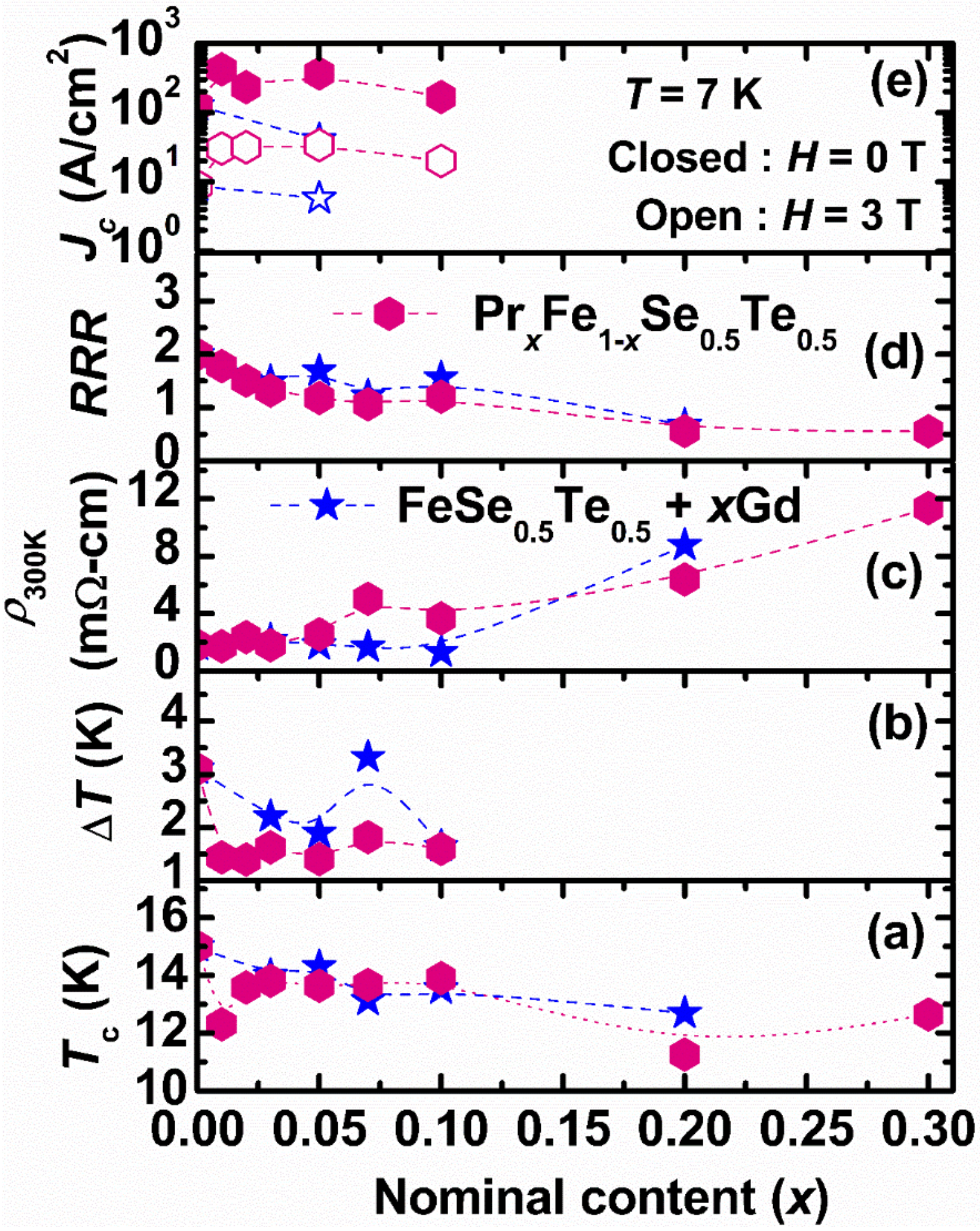

**Figure 11:** The variation of **(a)** the onset transition temperature ($T_c$) **(b)** the transition width ($\Delta T$) **(c)** the room temperature resistivity ($\rho_{300K}$) **(d)** residual resistivity ratio ($RRR = \rho_{300K} / \rho_{20K}$) **(e)** the critical current density ($J_c$) at 7 K for $H$ = 0T (**closed symbol**) and 3 T (**open symbol**) for $Pr_xFe_{1-x}Se_{0.5}Te_{0.5}$ bulks prepared by HP-HTS with respect to the nominal contents ($x$) of Pr substitutions or Gd- additions.

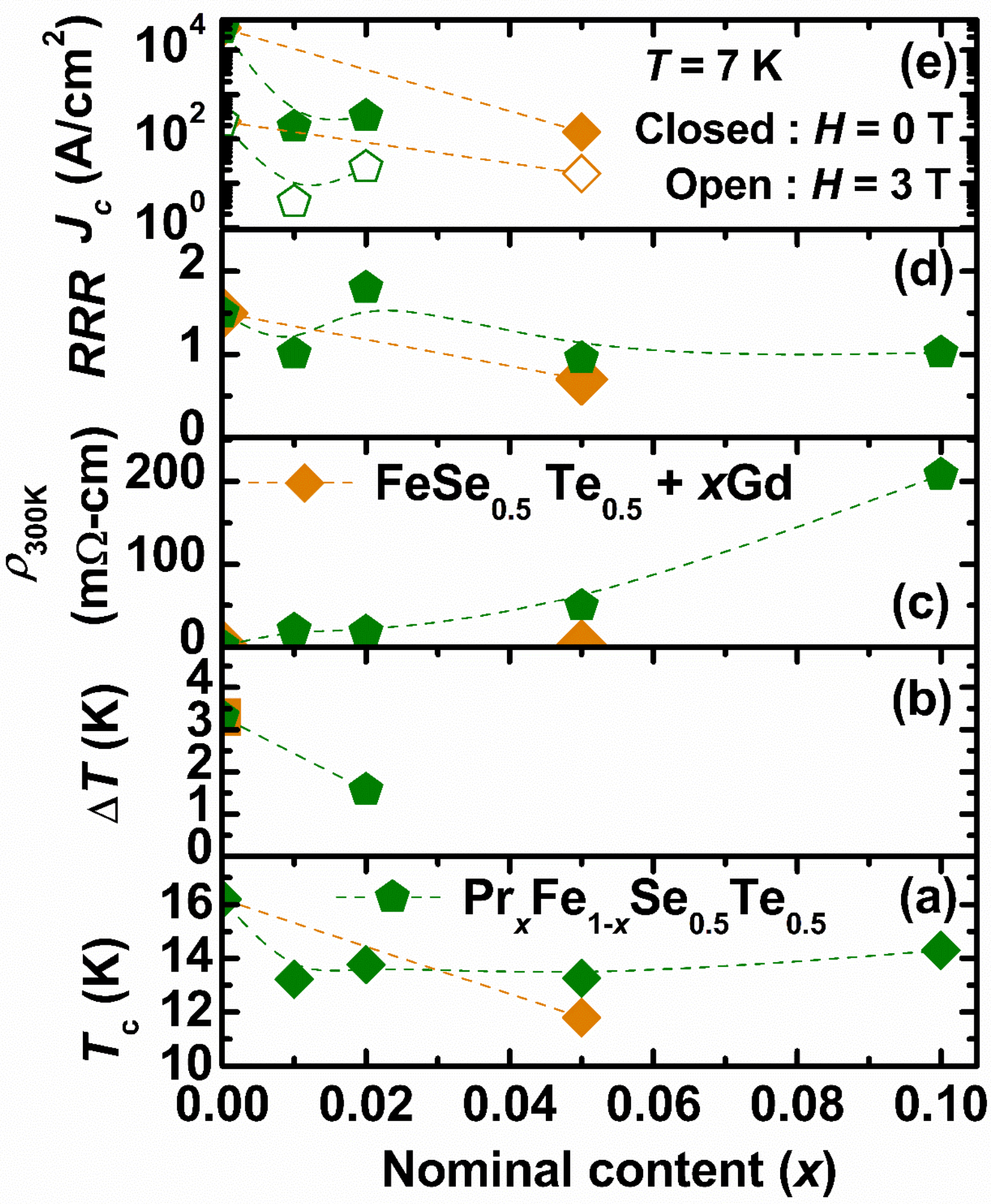